\documentclass[12pt]{article}
\usepackage{epsf}
\usepackage{amsmath}
\usepackage{graphicx}
\usepackage{color}
\usepackage{psfrag}
\usepackage{pdfpages}
\usepackage[draft]{todonotes}

\usepackage{epsf}
\usepackage{graphicx,epsfig}
\usepackage{amssymb}
\usepackage{epstopdf}
\usepackage{amsmath}
\usepackage{amsfonts}
\usepackage{latexsym}
\usepackage{color}
\usepackage{xypic}
\usepackage{cancel,slashed}
\usepackage[linktocpage]{hyperref}
\usepackage{cite}

\usepackage[latin1]{inputenc}
\usepackage{amsmath}
\usepackage{color}
\usepackage{colortbl}
\usepackage{amsfonts}
\usepackage{amssymb}
\usepackage{syntonly}
\usepackage{hyperref}
\usepackage{pdfpages}
\usepackage{dsfont}
\usepackage{graphicx}
\usepackage{caption}
\usepackage{subcaption}

\usepackage{ifpdf}

\def\s{{\text{sgn}}}

\def\mp {M_{\rm P}}
\def\vol {{\cal V}}

\def\im{\mbox{Im }}

\def\be{\begin{equation}}
\def\ee{\end{equation}}
\def\bea{\begin{eqnarray}}
\def\eea{\end{eqnarray}}
\def\bes{\begin{subequations}}
\def\ees{\end{subequations}}

\newcommand{\bmat}{\left(\begin{array}}
\newcommand{\emat}{\end{array}\right)}

\def\yzero{\smash{\hbox{$y\kern-4pt\raise1pt\hbox{${}^\circ$}$}}}

\def\beq{\begin{equation}}
\def\eeq{\end{equation}}
\def\beqa{\begin{eqnarray}}
\def\eeqa{\end{eqnarray}}

\def\-{\hphantom{-}}

\def\s2{\frac{1}{\sqrt2}}

\def\IF{\relax{\rm I\kern-.18em F}}
\def\II{\relax{\rm I\kern-.18em I}}

\def\Dsl{\,\raise.15ex\hbox{/}\mkern-13.5mu D} 
\def\id{{\bf  1}}

\def\im{{\rm Im}\,}

\catcode`\@=11   
\newdimen\@rotdimen
\newbox\@rotbox  

\def\@vspec#1{\special{ps:#1}}
\def\@rotstart#1{\@vspec{gsave currentpoint currentpoint translate
   #1 neg exch neg exch translate}}
\def\@rotfinish{\@vspec{currentpoint grestore moveto}}
\def\@rotr#1{\@rotdimen=\ht#1\advance\@rotdimen by\dp#1%
   \hbox to\@rotdimen{\hskip\ht#1\vbox to\wd#1{\@rotstart{90 rotate}%
   \box#1\vss}\hss}\@rotfinish}
\def\@rotl#1{\@rotdimen=\ht#1\advance\@rotdimen by\dp#1%
   \hbox to\@rotdimen{\vbox to\wd#1{\vskip\wd#1\@rotstart{270 rotate}%
   \box#1\vss}\hss}\@rotfinish}%
\def\@rotu#1{\@rotdimen=\ht#1\advance\@rotdimen by\dp#1%
   \hbox to\wd#1{\hskip\wd#1\vbox to\@rotdimen{\vskip\@rotdimen
   \@rotstart{-1 dup scale}\box#1\vss}\hss}\@rotfinish}%
\def\@rotf#1{\hbox to\wd#1{\hskip\wd#1\@rotstart{-1 1 scale}%
   \box#1\hss}\@rotfinish}%
\def\rotate{\@ifnextchar[{\@rotate}{\@rotate[l]}}
\def\@rotate[#1]#2{\setbox\@rotbox=\hbox{#2}\@nameuse{@rot#1}\@rotbox}

\catcode`\@=12

\topmargin
-1.5cm
\textwidth
15.5cm
\textheight
23.5cm
\oddsidemargin
0.7cm
\evensidemargin
1.2cm

\begin{document}

\makeatletter
\@addtoreset{equation}{section}
\makeatother
\renewcommand{\theequation}{\thesection.\arabic{equation}}
\renewcommand{\thefootnote}{\fnsymbol{footnote}}
\pagestyle{empty}
\rightline{IFT-UAM/CSIC-15-030}
\rightline{FTUAM-15-9}
\rightline{DESY-15-048}
\vspace{0.1cm}
\begin{center}
\LARGE{\bf Bifid Throats \\for Axion Monodromy Inflation  \\[12mm]}
\large{Ander Retolaza$^{1,2}$\footnote{ander.retolaza@uam.es},  Angel M. Uranga$^1$\footnote{angel.uranga@uam.es}, Alexander Westphal$^3$\footnote{alexander.westphal@desy.de}\\[3mm]}
\footnotesize{$^1$ Instituto de Fisica Te\'orica IFT-UAM/CSIC,\\[-0.3em] 
C/ Nicol\'as Cabrera 13-15, Universidad Aut\'onoma de Madrid, 28049 Madrid, Spain\\[2mm]
$^2$ Departamento de Fisica Te\'orica, Universidad Aut\'onoma de Madrid, 28049 Madrid, Spain\\[2mm]
$^3$ Deutsches Elektronen-Synchrotron, DESY, 22607 Hamburg, Germany}\\[9mm]

\small{\bf Abstract} \\[5mm]
\end{center}

\begin{center}
\begin{minipage}[h]{16.0cm}

We construct a simple explicit local geometry providing a `bifid throat' for 5-brane axion monodromy. A bifid throat is a throat that splits into two daughter throats in the IR, containing a homologous 2-cycle family reaching down into each daughter throat.  Our example consists of a deformed $\mathbb{Z}_3\times\mathbb{Z}_2$ orbifold of the conifold, which provides us with an explicit holographic dual of the bifid throat including D3-branes and fractional 5-branes at the toric singularities of our setup. Having the holographic description in terms of the dual gauge theory allows us to address the effect of 5-brane-antibrane pair backreaction including the warping effects. This leads to the size of the backreaction being small and controllable after imposing proper normalization of the inflaton potential and hence the warping scales.

\end{minipage}
\end{center}
\newpage
\setcounter{page}{1}
\pagestyle{plain}
\renewcommand{\thefootnote}{\arabic{footnote}}
\setcounter{footnote}{0}

\vspace*{1cm}

\setcounter{tocdepth}{2}

\tableofcontents

\vspace*{1cm}

\section{Introduction}

We are starting an era of precision $B$-mode observations in cosmology, especially since the recent results from BICEP2 and Planck  (see \cite{Ade:2015tva} for their combined analysis). Future observations \cite{Creminelli:2015oda}  will either detect or put stringent constraints on primordial $B$-modes from gravitational waves during inflation, therefore sharpening our picture of the very early universe, and providing new tools to discriminate among the plethora of present inflationary models/scenarios (see \cite{Baumann:2014nda} for a recent string-motivated review). Indeed, in single field inflation models, the Lyth bound \cite{Lyth:1996im} correlates the tensor to scalar ratio $r$ with the inflaton field range. Interestingly, the present observational bound $r<0.12$ is still compatible with large field inflation models, in which the field range is trans-Planckian and the inflation scale is very high.  Large-field inflation models
are sensitive to an infinite number of corrections to the inflaton potential which are suppressed by the Planck mass scale. The construction of viable models in a concrete framework of quantum gravity, such as string theory, is proving an interesting adventure. 

A natural way to suppress the couplings of the inflaton to the heavy degrees of freedom is through axions, i.e. periodic scalars with an approximate continuous shift symmetry. In string theory, there are two broad proposals to realize large-field inflation with axions \cite{Baumann:2014nda}. The first involves multiple axions \cite{Kim:2004rp,Dimopoulos:2005ac,Grimm:2007hs,Berg:2009tg,Ben-Dayan:2014lca, Ben-Dayan:2014zsa}, while the second utilizes a single axion with a non-trivial monodromy in field space (either arising via brane couplings\cite{Silverstein:2008sg,McAllister:2008hb} or via potentials from flux backgrounds \cite{Marchesano:2014mla,Blumenhagen:2014gta,Hebecker:2014eua,McAllister:2014mpa}, see also \cite{Garcia-Etxebarria:2014wla,Ibanez:2014kia,Franco:2014hsa,Blumenhagen:2014nba,Hebecker:2014kva,Ibanez:2014swa
}), see  \cite{Kaloper:2008fb,Kaloper:2011jz,Kaloper:2014zba} for a 4d phenomenological approach. In both cases, a trans-Planckian inflaton range  is compatible with a sub-Planckian  axion decay constant, in agreement with string theory expectations \cite{Banks:2003sx}.

The axion monodromy idea is particularly interesting in that the ingredients involved (shift symmetries, branes, possibly antibranes, and fluxes) are rather common in string theory\footnote{In addition, they seem to be free from the recently considered constraints that the weak gravity conjecture \cite{ArkaniHamed:2006dz} may impose  on transplanckian axion models \cite{Rudelius:2015xta,Montero:2015ofa,Brown:2015iha,Bachlechner:2015qja,Hebecker:2015rya,Brown:2015lia}.}. However, the construction of concrete string theoretical models is non-trivial. In recent axion monodromy models based on fluxes \cite{Marchesano:2014mla,McAllister:2014mpa}, the simplicity of the setup has allowed for various developments on model building and moduli stabilization (see e.g. \cite{Garcia-Etxebarria:2014wla,Blumenhagen:2014gta,Hebecker:2014eua,Ibanez:2014kia,Franco:2014hsa,Blumenhagen:2014nba,Hebecker:2014kva,Ibanez:2014swa}). On the other hand, the original models, based on supersymmetry breaking brane configurations \cite{Silverstein:2008sg,McAllister:2008hb} (see also \cite{Palti:2014kza}), require complicated geometries with multiple warped throats \cite{Flauger:2009ab}, which have not been amenable to detailed study. 

\begin{figure}[htb]
\begin{center}
\includegraphics[scale=.3]{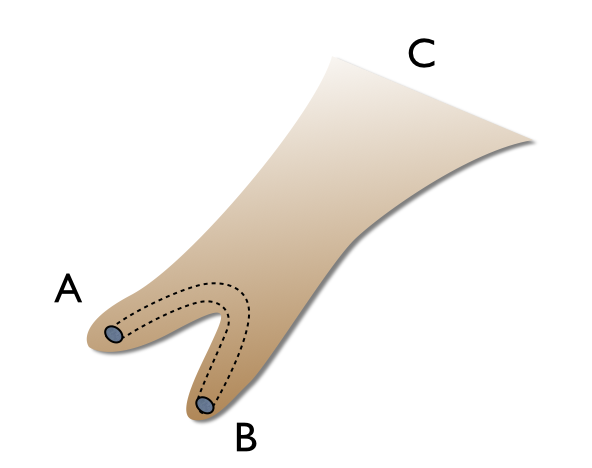}
\caption{\small Sketch of the bifid throat. The dashed line denotes the 3-chain showing that the two 2-cycles are homologous.}
  \label{fig:bifid}
\end{center}
\end{figure}

In more detail, the configurations in \cite{McAllister:2008hb,Flauger:2009ab} take the inflaton to be an axion coming from the  type IIB RR 2-form integrated over a 2-cycle. Actually, the geometry must contain two 2-cycles in the same homology class but located at the bottom of two different warped throats. Wrapping an NS5-brane and an NS-antibrane on these two 2-cycles, their charges cancel but their couplings to the RR axion add up, endowing it with a monodromic potential suitable to host large field inflation. The energy increase is associated to the appearance of induced D3 brane-antibrane  charge due to the axion shift. Finally, in order to suppress the backreaction of the NS brane-antibrane pair on modes localized on a complex dimension one region \cite{Conlon:2011qp}, the configuration must be located at the bottom of a common throat  \cite{Flauger:2009ab}. Such a geometry, which we dub {\em bifid throat}, is illustrated in Figure \ref{fig:bifid}.

The potential appeal of these models is concealed by the naive complexity of the underlying geometry. Actually, as we will show, relatively simple geometries can enjoy the right topological properties to host such systems. We work out the simplest such explicit example, based on a orbifold of the conifold; similar more involved examples could be worked out with the same techniques. The main motivation of this paper is to moreover provide a new handle on the bifid throat geometries required for brane-antibrane axion monodromy inflation. This is done  by providing a holographic dual field theory for the throat geometry (and to some extent, of the brane-antibrane system), generalizing the Klebanov-Strassler throat \cite{Klebanov:2000hb}. This allows a holographic description of the backreaction and of its suppression. Along the way, we provide a direct link between the microscopic description of the system and the effective field theory in \cite{Kaloper:2008fb,Kaloper:2011jz,Kaloper:2014zba}. We expect that this analysis improves the understanding of the  axion monodromy models from brane-antibrane pairs in throats, and allows to strengthen their formulation for model building in a near future full of observational promises.

The paper is organized as follows. In Section \ref{sec:effective} we frame the NS brane-antibrane models in the 4d effective description of \cite{Kaloper:2008fb,Kaloper:2011jz,Kaloper:2014zba}. This is not necessary for the rest of the paper, but is a complementary step to put these models on an even firmer ground. In Section \ref{sec:warmup} we provide the holographic field theory description of a simplified bifid throat, which contains all ingredients except the homologous 2-cycles at the infrared ends of the geometry. In section \ref{sec:dimer-simple}, the dual gauge theory is described as a system of D3-branes at a toric singularity, encoded in terms of dimer diagrams. Section \ref{sec:AganagicEtAl} reviews the construction of~\cite{Aganagic:2006ex} which allows to directly get a homologous 2-cycle family, and discusses the crucial difference our setup has by providing for independent holographic descriptions of several conifolds. Section \ref{sec:flow-simple} describes the deviation from conformality, by the introduction of fractional branes (triggering duality cascades dual to the different warped throats) and of Higgsing vevs (splitting the infrared geometry into two independent throats). In Section \ref{sec:real-thing} we perform a similar analysis for a bifid throat with the required homologous 2-cycles, and therefore realistic to produce axion monodromy. In section \ref{sec:dimer-real} we describe the holographic dual field theory using dimer diagrams, and describe the non-trivial RG flow in sections
\ref{sec:deform-real}, \ref{sec:elong-real}, \ref{sec:last-real}. In section \ref{sec:back-cascade} we describe the holographic view of introducing D5 brane-antibrane pairs, and describe their backreaction in field theory language, assessing it is localized at the energies of the neck connecting the infrared throats. A similar result is plausible for the S-dual configuration with NS branes. Finally,  section \ref{sec:conclusions} contains our conclusions.

\section{Effective lagrangian description}
\label{sec:effective}

Axion monodromy inflation was introduced in \cite{Silverstein:2008sg}, and concrete string constructions were proposed in \cite{McAllister:2008hb,Flauger:2009ab}, based on brane-antibrane pairs. A 4d effective field theory description of axion monodromy was suggested in \cite{Kaloper:2008fb} (see also \cite{Kaloper:2011jz,Kaloper:2014zba}). In \cite{Marchesano:2014mla}, this 4d description was  shown to appear in F-term axion monodromy models, where moreover the protection against UV corrections was shown to arise of the exact gauge invariance of a dual 3-form potential. In this section we show that the original models in \cite{McAllister:2008hb,Flauger:2009ab} can also be described in the 4d effective theory language, and their UV stability is thus also linked to an underlying gauge invariance. The discussion in this section is not necessary for the rest of the paper, (so the uninterested reader may skip it), but it provides a complementary step to base these models on an  even stronger ground.

Consider the type IIB brane-antibrane realization of axion monodromy, which involves an NS5-brane wrapped on a 2-cycle $\Sigma$, coupling to the axion as
\beqa
\phi=\int_{\Sigma} C_2
\eeqa
with $C_2$ the RR 2-form. The antibrane has a coupling of exactly the same kind, so we focus just on one.

The 4d effective description requires a coupling $\phi F_4$, for $F_4=dc_3$ the field strength of a suitable 3-form. In our case, the 3-form $c_3$ is the  6d worldvolume dual of the 1-form gauge field $A_1$ (i.e. ${*_{6d}}\, dc_3=F_2$). To show it, we use a chain of dualities to display the mixed coupling $\phi F_4$, as follows. Start with a D3-brane, whose worldvolume gauge field $A_1$ couples to the RR 2-form $C_2$ via
\beqa
\int_{\rm D3}C_2 \wedge F_2 \ .
\eeqa
Under an S-duality,  we get a similar D3-brane coupling of the  NSNS 2-form $B_2$ and the dual gauge field strength ${\tilde F}_2$. Performing two T-dualities, we turn the D3- into a D5-brane, while  $B_2$ remains invariant, and ${\tilde F}_2$ turns into the field strength $F_4=dc_3$ of the D5-brane magnetic dual gauge potential 
\beqa
\int_{\rm D5} B_2 \wedge F_4 \ .
\eeqa 
A final S-duality turns the D5 into an NS5, and also changes $B_2$ for $C_2$, to give a similar coupling 
with $F_4$ being the dual field strength of the NS5 brane worldvolume gauge field. Compactification on the 2-cycle $\Sigma$ yields the mixed term $\phi F_4$.

This mixed term is precisely that in the description in \cite{Kaloper:2008fb} (see also \cite{Kaloper:2011jz}), where a 4d 3-form $c_3$ coupling to the axion via a Lagrangian schematically 
\beqa
\int |F_4|^2 + \int |d\phi|^2 + \int \phi F_4
\label{ks-lagrangian}
\eeqa
with $F_4=dc_3$. The $|F_4|^2 $ term comes from the S-dual of the DBI action for the NS5-brane in the small field-strength approximation. Higher powers of the field strength arising along the same duality chain from the full DBI action, which should accordingly resum into a square-root expression, will flatten the initially quadratical potential for $\phi$ into the known linear regime at large field values.

Upon dualization of the periodic scalar into a 4d 2-form $b_2$, the action can be written
\beqa
\int_{4d}\, |db_2-c_3|^2 \ .
\eeqa
This theory has a gauge invariance 
\beqa
 c_3\to c_3+d\Lambda_2\quad , \quad b_2\to b_2+\Lambda_2 \ .
\eeqa
The above lagrangian describes the generation of an axion potential (and hence a mass term) in terms of a 3-form eating up a 2-form and becoming massive, in a higher $p$-form analogue of a Higgs mechanism. As emphasized in \cite{Marchesano:2014mla}, this underlying gauge symmetry constrains possible corrections, and ensures the viability of the axion monodromy potential for inflationary purposes.

\section{A simple bifid throat} 
\label{sec:warmup}


In this section we describe a simple geometry with the right ingredients to support two small throats (denoted the IR throats) at the bottom of a common one (the UV throat), and provide its holographic dual gauge theory. It arises as the worldvolume theory on a stack of D3-branes at the tip of a toric CY singularity, in the presence of fractional branes. As in \cite{Klebanov:2000hb}, the throats are dual to energy regimes in which the theory experiences cascades of Seiberg dualities, whereas the end of the throat is mapped to confining gauge dynamics and quantum deformations of the moduli space. In addition, the separation between the two infrared throats is dual to a Higgs mechanism induced by {\em classical} mesonic vevs (i.e. not arising from strong dynamics). The ordering (or relative geometry) of the throats is associated to the scales of confining dynamics and of the Higgsing.

The simple model in this section has all ingredients, except for the requirement of having a homologous 2-cycle on the two IR throats, recall Figure \ref{fig:bifid}. This latter property will be achieved in section \ref{sec:real-thing}, by simply adding an extra $\mathbb{Z}_2$ orbifold to the model in this section, which is therefore an optimal warmup exercise.
\vspace*{-6pt}
\subsection{The geometry and dual gauge theory}
\label{sec:dimer-simple}

As just mentioned, in this section we skip the requirement of having the 2-cycle at the end of the throat. 

We need a local geometry with three independent complex deformations, so that it contains three independent 3-cycles which support the fluxes producing the UV and the two IR throats. The problem of characterizing the complex deformations of a local CY singularity is in general difficult, but it has a simple answer for toric singularities. The criterion for a toric singularity to admit a complex deformation was discussed in \cite{Franco:2005fd}: the web diagram should admit a split into subwebs. We will consider a singularity which admits the removal of three independent subwebs to account for the three throats. The question of why two are inside a common one is a question of scales, as will be clear later on.

\begin{figure}[h]
\begin{center}
\includegraphics[scale=0.4]{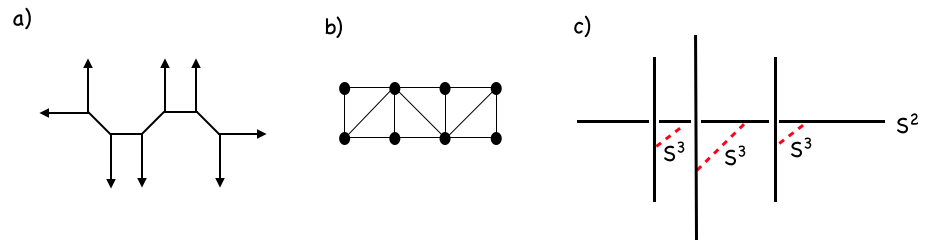} 
\caption{(a) Web diagram of the singularity of interest; for the sake of clarity we have depicted the collapsed 2-cycles of finite size. (b) Toric diagram, where the initiated easily recognizes an orbifold of the conifold. (c) Splitting of the web diagram displaying the three complex deformations of the geometry, and the three corresponding 3-cycles.}
\label{webdiagram}
\end{center}
\end{figure}

We will provide one explicit model, based on the simplest toric singularity with the desired properties; it is straightforward to construct other toric examples. The web diagram is shown in Figure \ref{webdiagram}(a), its dual toric diagram in (b) and the result of complex deformation is shown in (c). Each complex deformation is locally identical to a conifold transition, hence the 3-cycles are non-intersecting 3-spheres, which we denote by $A_{\rm UV}$, $A_{\rm IR,1}$, $A_{\rm IR,2}$. Their (non-compact) dual 3-cycles are denoted by $B$'s.

The physics of the throat can be very explicitly discussed in terms of the holographic dual gauge theory (with fractional branes). The gauge theory is that corresponding to D3-branes at the singularity of Figure \ref{webdiagram}(a), in the limit of collapsed 2-cycles. Since the singularity is toric, we can exploit the powerful tools of dimer diagrams (see e.g. \cite{Franco:2005rj,Kennaway:2007tq}) to construct the gauge theory. 

In fact, this geometry is easily recognized as a $\mathbb{Z}_3$ orbifold of the conifold. For completeness we provide its description. Describing the conifold by $xy-zw=0$, the $\mathbb{Z}_3$ orbifold is given by the action
\beqa
x\; \to\; e^{2\pi i/3} x\quad ; \quad y\;\to \; e^{-2\pi i/3} y \quad ;\quad x,y \; {\rm invariant} \ .
\eeqa
Defining the invariant coordinates $x'=x^3$, $y'=y^3$, the resulting space can be described by 
\beqa
x'y'-z^3w^3=0 \ .
\label{geom1}
\eeqa
It is easy to describe the three complex deformations. To do so, rewrite (\ref{geom1}) as $xy-t^3=0$, $zw=t$, and deform with three complex parameteres $\epsilon_i$, $i=1,2,3$ to
\beqa
&& xy=(t-\epsilon_1)(t-\epsilon_2)(t-\epsilon_3)\nonumber \\
&& zw=t \ .
\label{deform-simple}
\eeqa

\begin{figure}[htb]
\begin{center}
\includegraphics[scale=0.45]{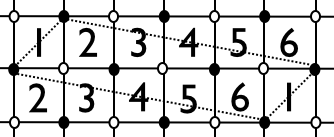} 
\caption{\small The dimer describing the gauge theory for the underlying system of D3-branes at the singular geometry in Figure \ref{webdiagram}. It corresponds to enlarging the unit cell in the infinite periodic array corresponding to the conifold dimer.}
\label{fig:dimer1}
\end{center}
\end{figure}

The dimer describing the field theory on a probe D3-brane on this throat is just that of the conifold with an order-3 enlargement of the unit cell, as shown in Figure \ref{fig:dimer1}.

\begin{figure}[h]
\begin{center}
\includegraphics[scale=0.45]{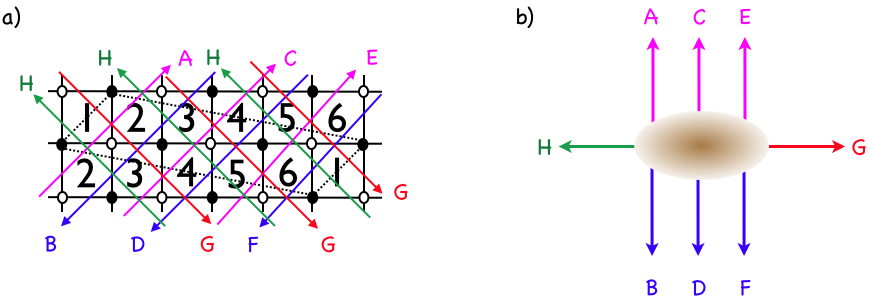}  
\caption{\small Zig-zag paths in the dimer, and picture of the external legs in the web diagram as obtained from their $(p,q)$-classes after a $SL(2,\mathbb{Z})$ transformation.}
\label{initialzigzag}
\end{center}
\end{figure}

One way to show that the dimer corresponds to the geometry of interest, is to draw the zig-zag paths and check their $(p,q)$ homology class in the $\mathbb{T}^2$ unit cell of the dimer. They  define the directions of the external legs of the web diagram for the geometry, as shown in Figure \ref{initialzigzag}.

\subsection{Comparison with the meta-stable SUSY breaking multi-conifold proposal in Aganagic et al.~\cite{Aganagic:2006ex}}
\label{sec:AganagicEtAl}

Aganagic et al.~\cite{Aganagic:2006ex} have provided another construction of meta-stable SUSY breaking via multiple conifolds with either 5- and anti-5-branes (the resolved phase) or 3-form fluxes of both signs (the deformed phase). We  shortly review their analysis here, because a comparison will show why we are using the $\mathbb{Z}_3$-orbifold setup discussed in the previous section.

Aganagic et al. start by describing an ${\cal N}=1$ gauge theory on a stack of $N$ D5-branes wrapped on a singular conifold with shrunken 2-cycle in a non-compact CY. The underlying conifold geometry is ${\cal N}=2$, which is broken by D5-stacks to one ${\cal N}=1$ and by anti-D5-stacks to a different orthogonal ${\cal N}=1$. The combined presence of branes and anti-branes, or in the dual deformed geometry, flux and anti-flux, breaks SUSY to ${\cal N}=0$.
\newpage
\begin{figure}[h]
\begin{center}
\includegraphics[scale=0.44]{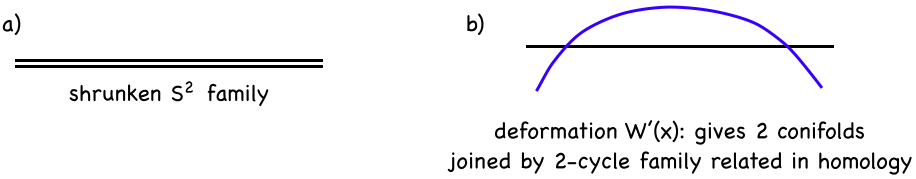}  
\caption{\small (a) Web diagram for the inital curves-of-conifolds geometry of Aganagic et al.~\cite{Aganagic:2006ex}. (b) The deformation $W'(x)$ leading to two conifolds connected by a 2-cycle family related in homology.}
   \vspace*{-15pt}
   \label{Aganagic1}
\end{center}
\end{figure}

\begin{figure}[t]
\begin{center}
\includegraphics[scale=0.44]{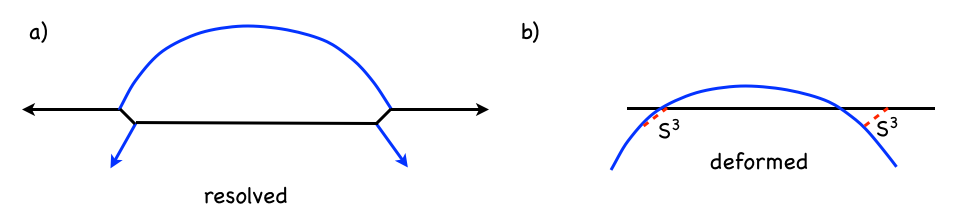}  
\caption{\small (a) Web diagram for the resolved 2-conifold geometry of Aganagic et al.~\cite{Aganagic:2006ex}. (b) The deformed version of (a) after the geometric transition.}
   \vspace*{-15pt}
\label{Aganagic2}
\end{center}
\end{figure}

The geometry of the setup starts from a \emph{curve} of $A_1$-singularities (conifold) described by the CY equation
\beq
uv=y^2
\eeq
on $(u,v,y,x)\in \mathbb{C}^4$. The locus $u=v=y=0$ describes an $A_1$ singularity at every $x$. If we resolve this, we get an entire curve's worth of resolved conifolds, and hence a whole family of holomorphic 2-cycles related in homology, see Figure~\ref{Aganagic1}(a). To get a set of individual conifolds, we add a deformation to the defining CY equation, which is chosen such that it restores the conifold condition only at $m$ different points $a_k$ in $x$. We can do this by writing
\beq
uv=y^2+W'(x)^2\quad,\quad W'(x)=g\prod\limits_{k=1\ldots m}(x-a_k) \ .
\eeq
Adding this polynomial in $x$ deforms the original CY away from the conifold locus everywhere except at the $m$ points $x=a_k$. If we now resolve this system, we get a family of 2-cycles, which are all related in homology and have $m$ holomorphic (minimum volume) representatives at the $m$ former conifold points. This can be seen in their volume expression
\beq
A(x)=\sqrt{r^2+W'(x)^2}
\eeq
where $r$ is the modulus of the resolution parameter. If we choose $m=2$ this system describes a setup of two resolved conifolds which are connected by a non-minimal volume family of 2-cyles in homology with a holomorphic representative at each end of the family. This is one way of realizing the 2-cycle-family geometry necessary to generate 5-brane axion monodromy. Indeed, according to~\cite{Aganagic:2006ex}, if we now wrap $N$ 5-branes magnetized by $B_2$ on the holomorphic 2-cycle on one end of this family, and $N$ magnetized anti-5-branes on the opposite one, the system generates meta-stable SUSY breaking. The meta-stability is visible geometrically, in that the expanding volume of the 2-cycle family between the two holomorphic representatives causes the 5-branes to require additional energy to cross their distance and annihilate. This happens because the branes need to expand when moved along the 2-cycle family with its increasing 2-cycle volume away from the resolved conifold points. Aganagic et al. show that the same effect follows after a geometric transition from a system where the two original conifold loci are deformed (instead of resolved, see Figure~\ref{Aganagic2})
\beq\label{eq:AganagicW2}
uv=y^2+W'(x)^2+f_{m-1}(x) \ .
\eeq
Here $f_{m-1}(x)$ denotes the deformation polynomial.
After this geometric transition the dual of the brane gauge theory is described by a flux superpotential \cite{Gukov:1999ya}
\beq
W=\int G_3\wedge \Omega=\alpha\,S+N\partial_S{\cal F}\ .
\eeq
Here we have $N$ units of $F_3$ 3-form flux replacing the $N$ D5-branes, and $-N$ units of $F_3$ 3-form anti-flux replacing the $N$ anti-D5-branes, and $\alpha$ units of $H_3$-flux which are the dual of the $B_2$ magnetization on the 5-branes. $S$ denotes the $A$-type period or 3-cycle of the two deformed  conifolds.

In the presence of both flux (5-branes) and anti-flux (anti-5-brane) the system has no supersymmetric ground state any more. Instead there exists a non-supersymmetric critical point at $\alpha+N\bar\tau=0$ where $\tau=\partial_{SS}{\cal F}$ with vacuum energy
\beq
V=\frac{2i}{\tau-\bar\tau}|\alpha+N\tau|^2\sim N\,{\rm Im}\,\alpha\ .
\eeq
Since the geometric transition tells us that $\im\alpha=\int_B H_3=\int_{\partial B} B_2\equiv b$ in the D5-brane theory on the resolved side of the transition,\footnote{Here $B$ denotes the B-type 3-cycle connecting the two deformed conifolds, which becomes a 3-chain on the resolved side of the geometric transition.} we see that this analysis reproduces the linear 5-brane-anti-5-brane axion monodromy potential for the $B_2$ axion.

However, in looking at Figures~\ref{Aganagic1} and \ref{Aganagic2} we clearly see that this system does not allow for placing  deformed conifolds with different 3-form fluxes \emph{unrelated} in homology. Hence, while the above geometry is practically tailor-made to describe a family of homological 2-cycles with a meta-stable pair of a 5-brane and an anti-5-brane at each end, it is complicated for the description of warping at each end of the 2-cycle family in terms of a local well-controlled Klebanov-Strassler (KS)-like geometry.

This is why we opt for the geometry in Figure~\ref{webdiagram}, since it allows us to describe the generation of several independent conifolds unrelated in homology. This also enables a well-defined holographic description of each singularity a la  KS, thus including warping. The manageable price we pay for this advantage is the absence of a manifestly built-in family of homologous 2-cycles. However, we will see below that a branching of the geometry in Figure~\ref{webdiagram} over an additional $\mathbb{Z}_2$-orbifolding will provide with a 2-cycle family as well. We will see then, that the description of just that 2-cycle family (not the whole singularity) in terms of complex geometry will resemble the Aganagic et al. two-conifold system.

\vspace*{-8pt}
\subsection{The holographic flow}
\label{sec:flow-simple}

On the geometry side, we introduce RR 3-form fluxes in the 3-cycles obtained upon complex deformation of the geometry, and NSNS 3-form fluxes on their dual (non-compact) 3-cycles. We denote by $M$, $P_1$, $P_2$ the RR 3-form flux quanta over $A_{\rm UV}$, $A_{\rm IR,1}$, $A_{\rm IR,2}$; in addition, we denote by $N$ the RR 5-form flux along the base of the cones in the internal geometry  (at some radial position, since it is sourced by the RR fluxes and changes in the radial direction).

In the dual gauge theory, the fluxes correspond to the introduction of fractional D-branes in the above singular CY. They just correspond to anomaly-free rank assignments in the dimer gauge theory in Figure \ref{fig:dimer1}. Since the theory is non-chiral, any assignment is allowed. As will be clear from the analysis below, we take the following rank assignment for the different gauge groups to match the holographic dual:
\beqa
n_2=N+P_1\quad , \quad  n_4=N+M \quad , \quad n_6=N+P_2 \quad , \quad  n_1=n_3=n_5=N
\label{ranksuv}
\eeqa
(clearly, due to the cyclic symmetry of the gauge theory, any cyclic permutation of the above rank assignment leads to the same results, up to relabeling). 
We assume that $N\gg M\gg P_1 ,P_2\gg 1$, in order to produce long throats in the dual, describable in the geometric regime, and such that $M$ corresponds to the UV throat and $P_1,P_2$ to the IR throats.

In addition to the above rank assignments, we must specify some vevs to trigger the symmetry breaking effects, to split the bottom of the UV throat into two IR throats, which are easier to specify later on.

\medskip

{\bf The UV cascade}

\begin{figure}[h]
\begin{center}
\includegraphics[scale=0.42]{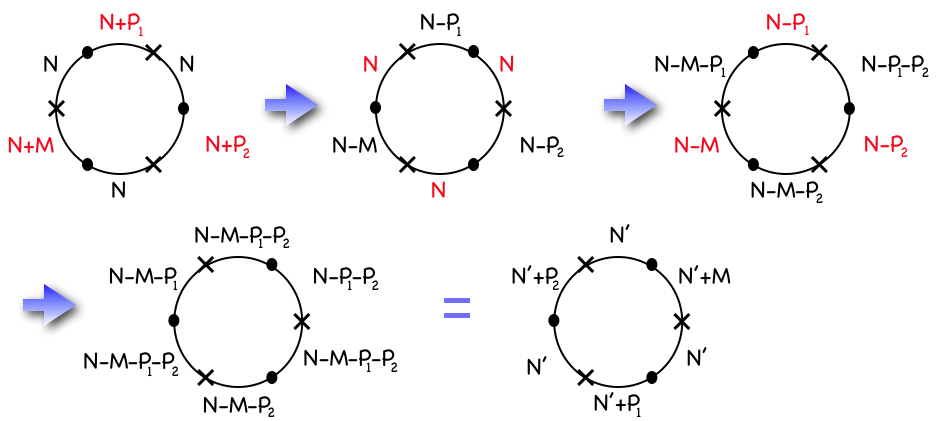}  
\vspace*{-5pt}
  \caption{\footnotesize Basic period of the UV cascade, in terms of a Hanany-Witten T-dual configuration of branes. Dots and crosses denote NS and NS'-branes localized on a periodic direction (denoted as the circle), with D4-branes suspended among them. The red labels denote the gauge factors experiencing Seiberg duality in going to the next step. Upon three such steps, one recovers a configuration identical to the original with the number of regular branes effectively decreased by $M+P_1+P_2$ (and a shift of the circle by 3 intervals).}
   \label{uvcascade}
\end{center}
\end{figure}

The dynamics starts at some UV scale with the above rank assignment. The RG flow takes the theory through a duality cascade, which reduces the effective value of $N$ as one runs to lower energies. The discussion of the Seiberg dualities involved in the cascade is most easily carried out in terms of a T-dual Hanany-Witten configuration \cite{Hanany:1996ie}, which for the present singularity was discussed in \cite{Uranga:1998vf}. Concretely, we T-dualize (\ref{geom1}) along the $\mathbb{S}^1$ parametrized by $\alpha$ in the orbit of
\vspace*{-4pt}
\beqa
x\to e^{i\alpha} x \quad ,\quad y\to e^{-i\alpha} y \ .
\label{orbit-one}
\eeqa
\vspace*{-4pt}
The degeneration locus of the $\mathbb{S}^1$ (namely, when $x=y=0$) corresponds to $z^3w^3=0$, and describes 3 NS branes at $z=0$ (and along $w$) and 3 NS-branes at $w=0$ (and along $z$). Changing to more standard Hanany-Witten brane configuration conventions, we obtain a set of three NS-branes (along the directions 012345) and three rotated NS-branes (along 012389, denoted by NS'-branes), and D4-branes (along 0123 and the periodic direction 6) suspended between them. The presence of the $M$ fractional branes triggers a set of dualities detailed in Figure \ref{uvcascade}, which essentially corresponds to a triple unfolding of the Klebanov-Strassler duality cascade in the conifold. It is however modulated by the presence of the $P_1$, $P_2$ fractional branes in the gauge factors 2, 6, such that the reduction in the number of regular D3-branes upon three steps in the duality cascade is $\Delta N=-(M+P_1+P_2)$.
\vspace*{-8pt}

\medskip

{\bf First complex deformation}

\begin{figure}[h]
  \begin{center}
  \includegraphics[scale=0.45]{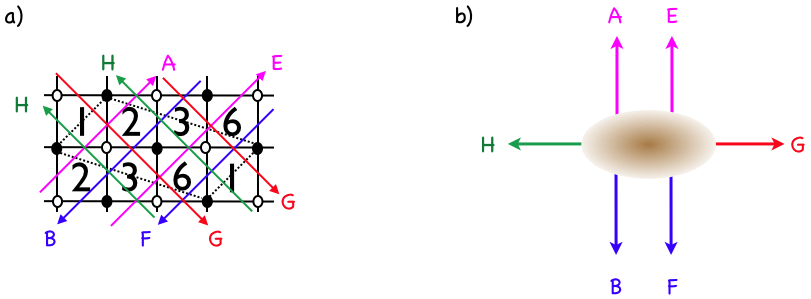}   
   \end{center}
  \caption{\small Result of the complex deformation of the initial geometry.}
  \label{deformation1}
\end{figure}

Let us start by taking $N=k(M+P_1+P_2)+M$. Then, after $k$ periods of the duality cascade we run out of D3-branes and reach the confinement regime dual to the complex deformation supporting the UV throat. Taking the last step, the ranks are as in (\ref{ranksuv}) with an effective value $N=M$, i.e.
\beqa
&& n_2=M+P_1\quad , \quad  n_4=2M \quad , \quad n_6=M+P_2 \nonumber \\
 && n_1=n_3=n_5=M \ .
\eeqa
For the gauge factor 4, we have $N_f=2N_c$ and there is a complex deformation of the moduli space, in analogy with  \cite{Klebanov:2000hb} (see also \cite{Franco:2005fd}). Accounting for the full non-perturbative dynamics of the gauge factors is easily done in terms of the dimer diagram \cite{Franco:2005fd}. Following this reference, the fractional brane corresponding to node 4 is a deformation brane associated to the removal of the legs C, D from the web diagram. The dimer diagram corresponding (or holographically dual) to the  left-over geometry after the deformation is obtained by removing the zig-zag paths C, D from the picture, and zipping together the unpaired remaining paths. This has the effect of recombining some of the faces, concretely 3 \& 5, that from now we will refer to as 3 (3 \& 5 $\rightarrow$ 3). Physically, this is because the mesons of the confined groups get vevs and this breaks part of the flavor symmetry.  The result of this operation is shown in Figure \ref{deformation1}.

The deformation is also easy to follow in the Hanany-Witten picture. It corresponds to the simultaneous removal of the NS and NS'-brane bounding interval 4, together with $M$ of the suspended D4-branes, hence recombining the intervals 3 and 5. 

In either picture, we are left with a $\mathbb{Z}_2$ quotient of the conifold, with nodes 1,2,3,6 and rank assignments 
\beqa
n_2=M+P_1\quad , \quad n_1=n_3=M\quad ,\quad n_6=M+P_2 \ .
\eeqa
It is possible to achieve a more general rank assignment, with the number of regular branes differing from $M$ (the fractional branes of the UV throat), by starting with $N=k(M+P_1+P_2)+M+Q$. The strong dynamics is trickier, and we simply quote that it leads to the same quantum deformation and $Q$ additional regular branes, in analogy with the Appendix in \cite{Klebanov:2000hb} for the conifold. In these cases, the strong dynamics typically corresponds to the appearance of a non-perturbative Affleck-Dine-Seiberg superpotential, whose F-term conditions enforce the quantum deformation of the moduli space of the left-over regular D-branes.

\medskip

{\bf Splitting the throat}

\begin{figure}[h]
\begin{center}
\includegraphics[scale=0.45]{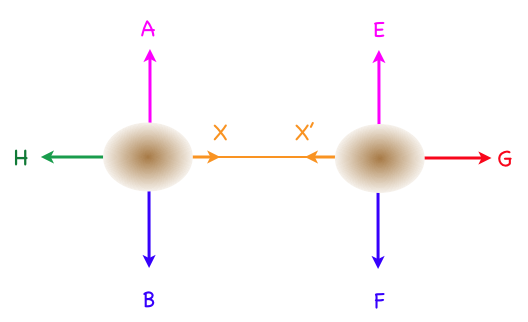}     
  \vspace*{-15pt}
 \caption{\small Elongation of the web diagram into two subwebs.}
\label{fig:elongation1}
\end{center}
\end{figure}

Once the first cascade has taken place, we reach a lower energy scale at which the gauge theory must split into two, corresponding to the two theories to become the duals of the two IR throats. Geometrically, the process is a splitting of the singularity into two remaining singularities, by a small resolution in which the web diagram is elongated (keeping it in the same plane) by separating the legs A,B,H from E,F,G. The  result is a factorization of the diagram into two,  one per left-over singularity (see Figure \ref{fig:elongation1}). At the level of the gauge theory, blowing up the singularity corresponds to turning on FI terms, whose contribution to the D-term potential must be cancelled by turning on suitable vevs, triggering a Higgs mechanism. Geometrically, fractional branes of the original singularity combine together to form fractional branes of the left-over singularities.

\begin{figure}[h]
\begin{center}
\includegraphics[scale=0.5]{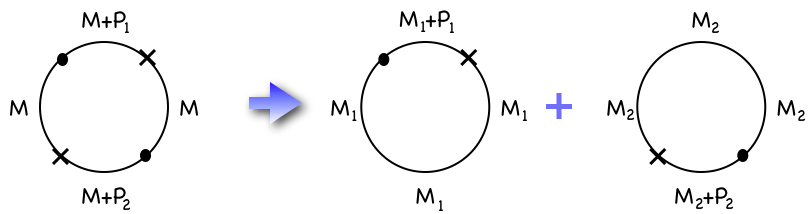}     
\caption{\small Higgs mechanism in the gauge theory in terms of a T-dual Hanany-Witten brane configuration. The partial blowup of the geometry, equivalent to the field theory FI terms, corresponds to the motion of one NS-NS' brane pair with respect to the other in the transverse direction 7, which enforces partitioning the stacks of D4-branes and recombining them across some intervals. The resulting two diagrams on the right-hand side are separated in the direction 7, and describe two decoupled conifold theories.}
 \label{fig:split1}
 \end{center}
\end{figure}

This can be easily reproduced using the Hanany-Witten brane configurations, as shown in Figure \ref{fig:split1}.

\begin{figure}[h]
 \begin{center}
 \includegraphics[scale=0.5]{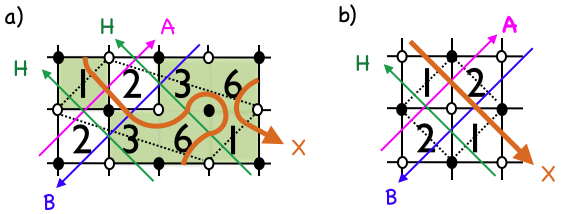}  
\caption{\small Dimer of the first gauge theory resulting from the elongation of the web diagram.}
  \label{conifold1zigzag}
 \end{center}
\end{figure}

The same result can be recovered using the tecniques in \cite{GarciaEtxebarria:2006aq}. Basically, the gauge theory splits into two, associated to the subsets (A,B,H) and (E,F,G). To get the first gauge theory sector, we draw the dimer diagram with only the zig-zag paths A,B,H and complete the unpaired paths by introducing a new one, labeled X. The edges not touched by A,B, H are precisely those bifundamentals getting a vev in the Higgssing. This breaks some of the gauge factors to their diagonal, specifically, 1,3,6 are combined together (and subsequently denoted by 1). The result of the operation is shown in Figure \ref{conifold1zigzag}(a), and simplified in (b), by a contraction of the diagram that corresponds to integrating out some massive fields. The resulting theory is  simply a conifold theory. 

\begin{figure}[h]
\begin{center}
\includegraphics[scale=0.5]{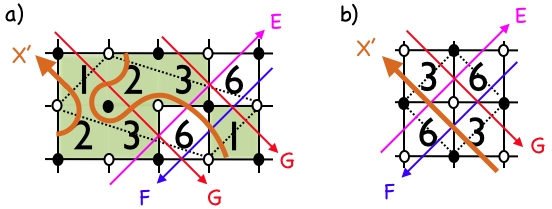}  
  \caption{\small Dimer of the second gauge theory resulting from the elongation of the web diagram.}
  \label{conifold2zigzag}
\end{center}
\end{figure}

In the same way, to get the second gauge theory, we draw the dimer diagram with only the zig-zag paths  E, F \& G, and complete them with a new path denoted X$'$. The process is a Higgs mechanism in which the gauge factors 1, 2,  3 are broken to the diagonal, subsequently denoted by 3. The resulting theory is shown in Figure  \ref{conifold2zigzag}, and corresponds to a second conifold theory.
\newpage

In purely field theoretical terms, the above operations in either brane picture correspond to turning on vevs of the form
\vspace*{3pt}
\beqa
\Phi_{23}=\Phi_{12}^T=\begin{pmatrix} 0_{(M_1+P_1)\times M_1} & \cr & v_2\,  \id_{M_2\times M_2} \end{pmatrix}
\nonumber\\[2pt]
\Phi_{61}=\Phi_{36}^T=\begin{pmatrix} v_1 \id_{M_1\times M_1}& \cr & 0_{ (M_2+P_2)\times M_2}  \end{pmatrix}  \ .
\label{vevs1}
\eeqa
\vspace*{3pt}
In words, the first matrix takes the $SU(M+P_1)$ theory at node 2, and breaks it with vevs for $M_2$ of its flavours $Q=\Phi_{23}$, ${\tilde Q}=\Phi_{12}$, breaking also the $SU(M)^2$ flavour symmetry. The surviving group is $SU(M_1+P_1)_2\times SU(M_2)_{123} \times SU(M_1)_1\times SU(M_1)_3$ \footnote{$SU(M_2)_{123}$ stands for the $SU(M_2)$  diagonal subgroup coming from gauge groups 1, 2 \& 3 after the Higgsing by (\ref{vevs1}). }. The second matrix corresponds to taking the $SU(M+P_2)$ theory at node 6, and giving mesonic vevs to $M_1$ of its flavours $Q=\Phi_{61}$, ${\tilde Q}=\Phi_{36}$, breaking the $SU(M)^2$ symmetry. The surviving group is $SU(M_1)_{136}\times SU(M_2+P_2)_6\times SU(M_2)_1\times SU(M_2)_3$. The actual symmetry surviving both Higgsings is $SU(M_1+P_1)_2\times SU(M_1)_{136}$ and $SU(M_2+P_2)_6\times SU(M_2)_{123}$, that we denoted $SU(M_1+P_1)_2\times SU(M_1)_{1}$ and $SU(M_2+P_2)_6\times SU(M_2)_{3}$ in the dimers of Figure \ref{conifold1zigzag} and Figure \ref{conifold2zigzag}. It is simple but tedious to check that the field theory Higgsing leads to two decoupled conifold gauge theory sectors, in agreement with the geometric splitting of  the D3-branes on the two left over conifold singularities.

\medskip

{\bf Smaller throats}

Below the scale of the symmetry breaking, the massive fields can be integrated out and we recover two decoupled conifold theories. Each independent conifold theory has fractional branes which can trigger standard Klebanov-Strassler throats \cite{Klebanov:2000hb}, providing the holographic dual of the two smaller throats. This part of the discussion is standard and requires no further comment.

\medskip

One last remark concerns the ordering of scales. The geometry of the throats corresponds to a precise ordering of the scales of strong dynamics for the UV gauge theory, $\Lambda$, the scale of symmetry breaking vevs $v$, and the strong dynamics scales of the final conifold theories $\Lambda_1$, $\Lambda_2$. Concretely, we need
\beqa
\Lambda\gg v\gg \Lambda_1,\Lambda_2 \ .
\eeqa
It is possible, but uninteresting for our purposes to consider other orderings, which would lead to different throat geometries.

\section{Bifid throat with homologous 2-cycles}
\label{sec:real-thing}

In this section, we construct a bifid throat similar to that in the previous section, but including  homologous 2-cycles at the tip of the IR throats. The simplest way to achieve this is to  consider a $\mathbb{Z}_2$ orbifold of the geometry in the previous section (hence a $\mathbb{Z}_3 \times \mathbb{Z}_2 $ orbifold of the conifold). To be concrete, we quotient (\ref{geom1}) by the action $z\to -z$, $w\to -w$; defining the invariants $z'=z^2$, $w'=w^2$, $t'=zw$ we have
\beqa
x'y'=t'{}^3 \quad , \quad z'w'=t'{}^2  \ .
\label{true-singu}
\eeqa
This produces a (complex) curve of $\mathbb{C}^2 / \mathbb{Z}_2$ singularities (manifest in the second equation above), which we will show to fall inside both IR throats, and whose  blown-up 2-cycle provides the (common) homology class where  the brane-antibrane pair will ultimately wrap. In this section we focus on the construction of the geometry, and postpone the introduction of the branes to section \ref{sec:thepair}.

The construction, even after the inclusion of fractional branes dual triggering the complex deformations supporting the fluxes in the dual geometry, is simply a $\mathbb{Z}_2$ quotient of that in the last section. Although it does not admit a simple T-dual Hanany-Witten brane configuration, it remains toric and can be easily described using dimer diagrams, which are just given by a two-fold extension of the dimers in the previous section. We therefore keep our discussion sketchy, as most ideas should already be familiar.

The web diagram for the geometry is shown in Figure \ref{fig:web} (a), its toric diagram in (b) and the result of the complex deformations is shown in (c). The existence of a curve of $\mathbb{C}^2/\mathbb{Z}_2$ singularities, even after the complex deformations, is manifest in the presence of two sets of parallel horizontal legs in the web diagram.

\begin{figure}[htb]
\begin{center}
\includegraphics[scale=.42]{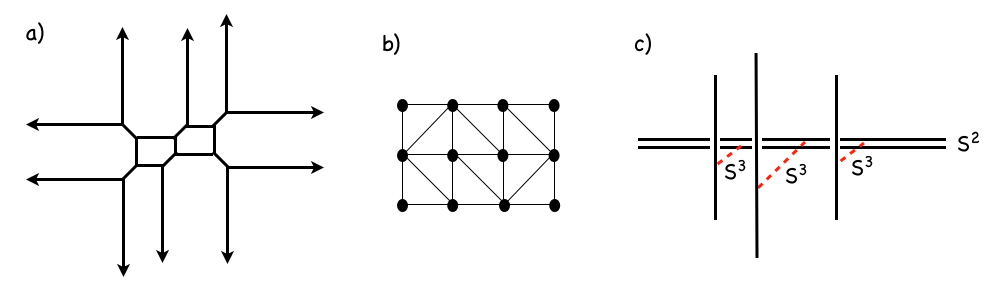}
\label{fig:web}
\caption{\small (a) Web diagram of the singularity of interest. (b) Its toric diagram. (c) Complex deformations of the geometry, showing the 3-cycles and the left-over curve of collapsed 2-spheres.} \label{fig:web}
\end{center}
\end{figure}

\subsection{The dimer}
\label{sec:dimer-real}

\begin{figure}[h]
\begin{center}
\includegraphics[scale=0.45]{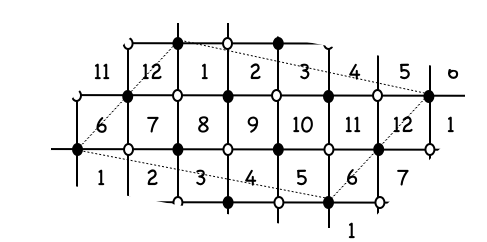}
\label{fig:dimer}
\caption{\small The dimer describing the gauge theory for the underlying system of D3-branes at the singular geometry.} 
\label{fig:dimerz6}
\end{center}
\end{figure}

As  previously done for the $\mathbb{Z}_3$ orbifold of the conifold, the dynamics of a D3-brane probing our geometry can be nicely encoded by using dimer diagrams. The dimer is shown in Figure \ref{fig:dimerz6}. To show that it corresponds to the geometry of interest, we draw the zig-zag paths and read their $(p,q)$ homology class in the $\mathbb{T}^2$ unit cell of the dimer, which define the directions of the external legs in the web diagram, corresponding to our geometry, see Figure \ref{fig:paths}.

\begin{figure}[t]
\begin{center}
\includegraphics[scale=.5]{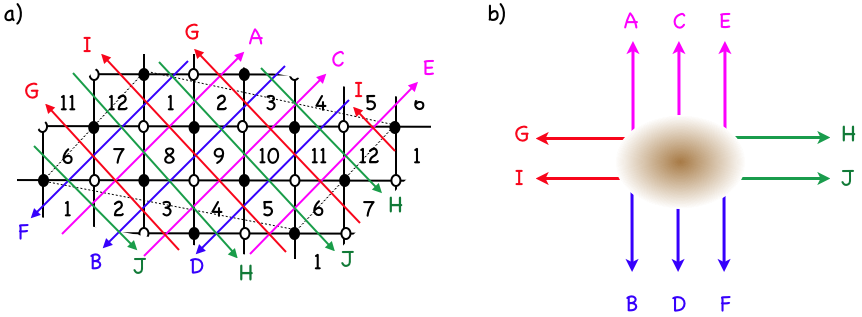}
\label{fig:paths}
\caption{\small Zig-zag paths in the dimer, and picture of the external legs in the web diagram as obtained from their $(p,q)$-classes.}
\label{fig:paths}
\end{center}
\end{figure}

According to the structure of our system, which is a $\mathbb{Z}_2$ orbifold of the construction in the previous section, our choice of fractional branes is $n_{i+6}=n_i$ with $n_i$ as in (\ref{ranksuv}), namely
\beqa
& n_2=n_8=N+P_1\quad , \quad  n_4=n_{10}=N+M \quad , \quad n_6=n_{12}=N+P_2 \nonumber \\ 
& n_1=n_3=n_5=n_7=n_{9}=n_{11}=N \ .
\label{ranksuvtwo}
\eeqa
The UV cascade proceeds as in section \ref{sec:flow-simple}, by simply operating on nodes $i$ and $i+6$ simultaneously. This preserves the $\mathbb{Z}_2$ symmetry throughout the process, so the dimer remains the two-fold extension of the dimers in the previous section, with the $n_{i+6}=n_i$ rank assignment rule. 

\subsection{First complex deformation: the common throat}
\label{sec:deform-real}

\begin{figure}[h]
\begin{center}
\includegraphics[scale=.5]{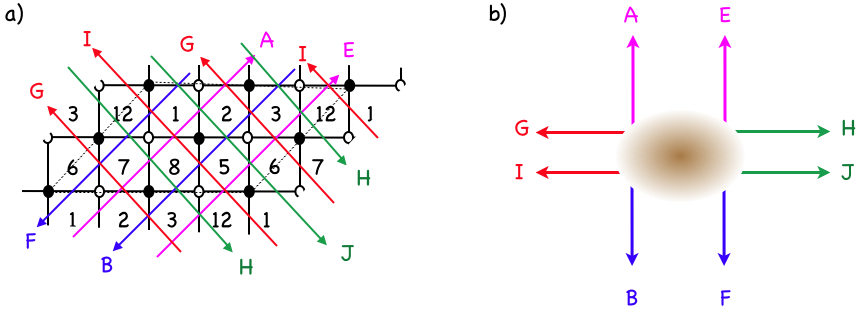}
\caption{\small Result of the complex deformation of the initial geometry.}
\label{fig:paths2}
\end{center} 
\end{figure}

\begin{figure}[h]
\begin{center}
\includegraphics[scale=.5]{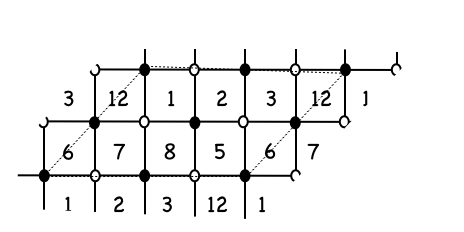}
\caption{\small Dimer of the gauge theory at the end of the first cascade.}
\label{fig:dimer2}
\end{center}
\end{figure}

As in section \ref{sec:flow-simple},  we eventually run out of regular D3-branes and encounter the complex deformation of the geometry. The complex deformation corresponding to the removal of the legs C, D from the web diagram, is triggered by the $M$ fractional branes on faces 4, 10 in the dimer (precisely those bounded by the paths C, D), see Figure \ref{fig:paths}. The gauge theory dynamics is (a two-fold extension) of that in the previous section, and the remaining field theory after the complex deformation is obtained  by similar diagrammatics. Namely, we remove the the paths C, D, and zip up unpaired paths. The gauge groups 5 and 9 are combined toghether (we label the result by 5), and so are 3 and 11 (labeled 3 henceforth). The result of this operation is shown in Figure \ref{fig:paths2}, and the dimer is displayed more cleanly in Figure \ref{fig:dimer2}. It corresponds to a $\mathbb{Z}_2\times \mathbb{Z}_2$ orbifold of the conifold. The remaining rank assignment is
\beqa
& n_2=n_8=M+P_1\quad , \quad  n_6=n_{12}=M+P_2 \nonumber \\ 
& n_1=n_3=n_5=n_7=M \ .
\label{ranks-higgs-two}
\eeqa

\subsection{Separating the stacks}
\label{sec:elong-real}

\begin{figure}[htb]
\begin{center}
\includegraphics[scale=.48]{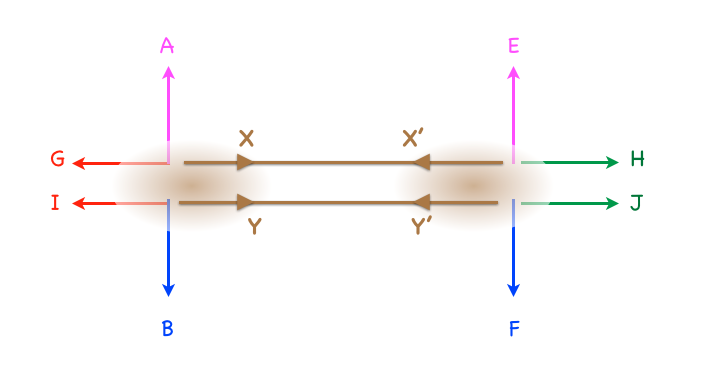}
\caption{\small Elongating the web diagram into two effective sub-singularities.}
\label{fig:elongate2}
\end{center}
\end{figure}

After the deformation / strong dynamics at the IR of the first throat/cascade, we reach a lower energy scale at which the gauge theory must split into two.  Geometrically, this is a resolution of the singularity in which the web diagram is elongated (keeping it in the same plane) by separating the legs A,B,G,I from E, F, H, J. The end result is a factorization of the diagram into two, see Figure \ref{fig:elongate2}.

At the level of the gauge theory, this corresponds to the introduction of FI terms, whose D-terms are cancelled by suitable bifundamental vevs, which Higgs down the gauge theory and split it in two sectors.
The field theory analysis is enormously simplified in terms of the dimer diagrams \cite{GarciaEtxebarria:2006aq}, as follows. To get the first gauge theory sector, we draw the dimer diagram with only the zig-zag paths A,B,G,I, and complete the unpaired paths with  new ones, in this case two, labeled X,Y. The edges not touched by A,B, G,I are precisely those bifundamentals getting a vev in the Higgsing.  

\newpage

\begin{figure}[htb]
\begin{flushleft}
\includegraphics[scale=.5]{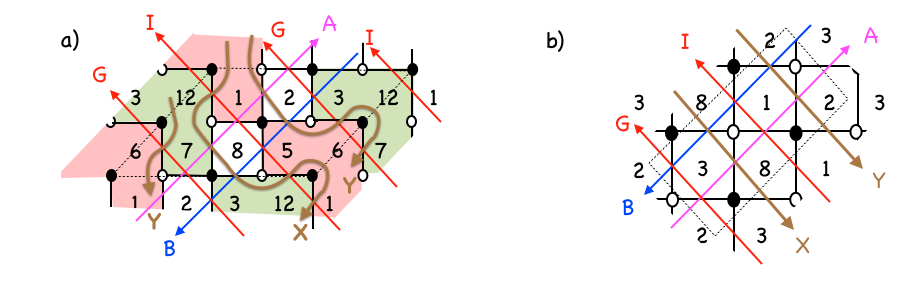}
\caption{\small Dimer of the first gauge theory resulting from the elongation of the web diagram.}
\label{fig:resolv1}
\end{flushleft}
\end{figure}

The diagrammatic process breaks some of the gauge factors to their diagonal, especifically, 3,7,12 are combined together (and subsequently denoted by 3), and so are 1,5,6 (herefrom denoted by 1). The result of the operation is shown in Figure \ref{fig:resolv1} (a), and simplified in (b), by a contraction of the diagram that corresponds to integrating out some massive fields. It corresponds to the dimer of a $\mathbb{Z}_2$ orbifold of the conifold (in agreement with the fact that its zig-zag paths reproduce, by construction, those of the web for such geometry). For concreteness, the orbifold action on $xy-zw=0$ is $z\to -z$, $w\to -w$, as inherited from the $\mathbb{Z}_2$ action at the beginning of section \ref{sec:real-thing}.

The rank assignments in this gauge theory sector are
\beqa
n_3=n_1=M_1 \quad , \quad n_2=n_8=M_1+P_1 \ .
\label{ranks-throat1}
\eeqa

\newpage

\begin{figure}[htb]
\begin{center}
\includegraphics[scale=.5]{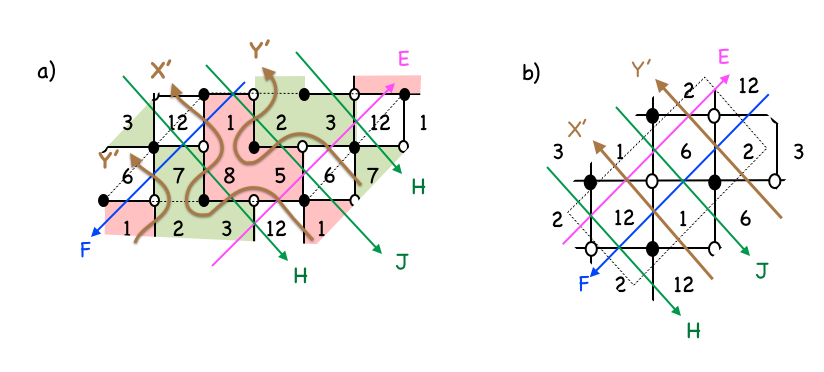}
\caption{\small Dimer of the second gauge theory resulting from the elongation of the web diagram.}
\label{fig:resolv2}
\end{center}
\end{figure}

The second gauge theory sector is obtained by drawing the dimer diagram with only the zig-zag paths E, F, H, J, and then completing the unpaired paths by two new ones, labeled X',Y'. The factors 2,3,7 are broken to the diagonal (denoted by 2), and so are 1,5,8 (denoted by 1 from now on). The operation is shown in Figure \ref{fig:resolv2}.

The rank assignments in this gauge theory sector are
\beqa
n_2=n_1=M_2 \quad , \quad n_6=n_{12}=M_2+P_2 \ .
\label{ranks-throat2}
\eeqa

The resulting geometries are two copies of a $\mathbb{Z}_2$ orbifold of the conifold. It is important to point out that both  small throats  pass through the same curve of $\mathbb{C}^2/\mathbb{Z}_2$ singularities, so both singularities share a common homology class for one of their 2-cycles. This is manifest from the web diagram, where the two parallel legs responsible for the $\mathbb{C}^2/\mathbb{Z}_2$ are common to both sub-diagrams.

At the field theory level, the explicit expression for the vevs can be simply obtained from the above dimer analysis, following \cite{GarciaEtxebarria:2006aq}, or by taking a two-fold extension of the result for the simpler construction in section \ref{sec:warmup}. For completeness we quote the result:
\vspace*{5pt}
\beqa
\Phi_{81}=\Phi_{58}^T=\Phi_{27}=\Phi_{32}^T=\begin{pmatrix} 0_{(M_1+P_1)\times M_1} & \cr & v_2\,  \id_{M_2\times M_2} \end{pmatrix}
\nonumber\\[3pt]
\Phi_{12,7}=\Phi_{3,12}^T= \Phi_{61}=\Phi_{56}^T=\begin{pmatrix} v_1 \id_{M_1\times M_1}& \cr & 0_{ (M_2+P_2)\times M_2}  \end{pmatrix} \ .
\label{vevs-complete1}
\eeqa
\vspace*{5pt}
\subsection{Last complex deformations: the small throats}
\label{sec:last-real}

 The rank assignments (\ref{ranks-throat1}), (\ref{ranks-throat1}) show that the two gauge theories are the holographic duals of configurations with $P_1$ and $P_2$ fractional branes, and hence define the UV of the subsequent duality cascades. Since the two sectors are very similar, we just discuss one of them. The cascade for this orbifold of the conifold has already been discussed in \cite{Franco:2005fd}, and merely corresponds to a two-fold extension of the Klebanov-Strassler conifold cascade. The IR physics is also similar, and leads to a quantum deformation of the moduli scape, dual to a complex deformation of the geometry, see below.
 
 The geometry is simple enough to be described explicitly. As advanced in the previous section, the $\mathbb{Z}_2$ orbifold action on the conifold
 \beqa
 xy-zw=0
 \eeqa
is defined by $x\to -x$, $y\to -y$. Introducing $x'=x^2$, $y'=y^2$, the orbifold of the conifold is
 \beqa
 x'y'-z^2w^2=0 \ .
 \eeqa
 There are two curves of $\mathbb{C}^2/\mathbb{Z}_2$ singularities at $x'=y'=z=0$ and $x'=y'=w=0$; in other words, at $x'=y'=0$ and $zw=0$.  The complex deformation is explicitly described by considering the same quotient but for the deformed conifold $xy-zw=\epsilon$, namely
 \beqa\label{eq:PseudoAganagic}
 x'y'=(zw-\epsilon)^2
 \eeqa
which clearly contains  $\mathbb{C}^2/\mathbb{Z}_2$ singularities (of the form $x'y'=t^2$) along the curve $x'=y'=0$ and $t\equiv zw - \epsilon=0\,$.\footnote{We note here, that the description of the deformation of just the two curves of $\mathbb{C}^2/\mathbb{Z}_2$ singularities sitting \emph{inside} our full construction in eq.~\eqref{eq:PseudoAganagic} has the same form as the deformed two-conifold geometry of~\cite{Aganagic:2006ex} given in eq.~\eqref{eq:AganagicW2} in section 3.2. It is therefore clear, that~\cite{Aganagic:2006ex} provided the description of the geometrically metastable 2-cycle family for which we gave here a full embedding into a toric singularity with KS-like duals.}

\medskip

\begin{figure}[htb]
\begin{center}
\includegraphics[scale=.5]{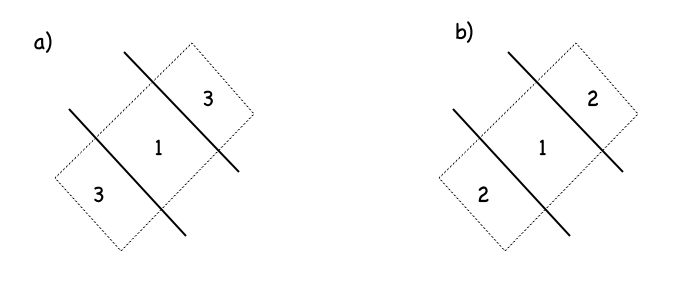}
\caption{\small Dimer of the gauge theories after the complex deformation at the bottom of the small throats.}
\label{fig:final}
\end{center}
\end{figure}

Let us carry out the gauge theory analysis in terms of the dimer diagrams. In the theory shown in Figure \ref{fig:resolv1} (b), the effect of the complex deformation corresponding to removing the legs A, B. Following \cite{GarciaEtxebarria:2006aq}, we remove the paths A, B from the dimer, and zip together the unpaired paths. The gauge factors 2 and 8, corresponding to  the fractional branes, disappear (due to confinement), and in this case nodes 1 and 3 remain independent. The remaining picture is shown in Figure \ref{fig:final}(a), and corresponds to a dimer associated to $\mathbb{C}^2/\mathbb{Z}_2$, as expected. A similar operation in the second gauge theory produces the picture in Figure \ref{fig:final}(b).

Notice that even though the two gauge theories are $\mathbb{C}^2/\mathbb{Z}_2$, by construction we are ensured that they belong to the same curve of singularities. Therefore, the 2-cycle in the blowup of this singularity falls inside both throats. This can be seen in the gauge theory language, because the fractional branes of the $\mathbb{C}^2/\mathbb{Z}_2$ in the first gauge theory are the same as those in the second (modulo gauge factors which have confined, i.e. whose homology 2-class has become trivial in the geometry). More explicitly, in the first $\mathbb{C}^2/\mathbb{Z}_2$ gauge theory, the final fractional branes correspond to labels 1, and 3. Now each of these came from the recombination of the original faces, specifically 1 comes from the set (1,5,6,9) and 3 comes from (3,7,11,12). Similarly, in the second gauge theory, the fractional branes carry labels 1 and 2, and actually correspond to the faces (1,5,8,9) and (2,3,7,11) of the original gauge theory. Since the faces 2,8,6,12 of the original theory have actually disappeared by confinement,  they do not define non-trivial homologly classes in the dual throat. Hence, the two fractional branes carry charges corresponding to the sets (1,5,9) and (3,7,11) in either of the two theories, consistently with the fact that they belong to the same curve of singularities.

\section{Inclusion of fivebrane-antibrane pair and axion monodromy}
\label{sec:thepair}

In this section we use the above holographic picture as a framework to study the brane-antibrane system wrapped on the homologous 2-cycle at the tip of the two final throats. As explained, these systems provide a realization of axion monodromy inflation \cite{McAllister:2008hb,Flauger:2009ab}. We hope that the holographic realization can provide interesting complementary views on these applications.

In our discussion, we consider the simpler setup of D5 brane-antibrane pairs wrapped on the 2-cycle. In applications to inflation, an NS5 brane-antibrane pair was proposed; this is because such branes couple to an axion coming from integrating the RR 2-form over the 2-cycle, and such axion scalar was argued not to appear in the K\"{a}hler potential in type IIB compactifications with O3/O7-planes. However, many local features of the system can be analyzed by considering the realization in terms of D5-branes (and performing S-duality if necessary). Moreover, D5-brane realizations may be interesting in their own right in global setups beyond O3/O7 CY compactifications. Hence we stick to the D5-brane picture in what follows.

The description of D5 branes (antibranes) wrapped on the 2-cycle corresponds to the inclusion of suitable fractional branes (antibranes) with respect to the $\mathbb{C}^2/\mathbb{Z}_2$, and is hence very simple. The 2-cycle is visible in the web diagram in Figure \ref{fig:elongate2} as the segment stretching between the legs $X$, $Y$. Then,  adding $K$ extra wrapped branes corresponds to increasing the ranks on the faces of the dimer enclosed by the corresponding zig-zag paths, for instance 1, 8, see Figure \ref{fig:resolv1}(b). The rank assignments change from (\ref{ranks-throat1}) to
\beqa
& n_1=M_1+K \quad , \quad n_8=M_1+P_1+K\nonumber \\
& n_3=M_1 \quad , \quad n_2=M_1+P_1 \ .
\label{ranks-throat1-extra}
\eeqa
The addition of the extra $K$ branes has a small backreaction on the RG cascade, which will be described in some more detail in section \ref{sec:back-cascade}.

It is convenient to trace this rank change up in the UV to the theory before the Higgsing. The addition of the $K$ fractional branes corresponds to a modification of the ranks on faces bound by paths $X$, $Y$, see Figure \ref{fig:resolv1}(a), namely 1,5,6,8. The rank assignments change from (\ref{ranks-higgs-two}) by
\beqa
& & \Delta n_1=\Delta n_{5}= \Delta n_6=\Delta n_{8}=K \ .
\label{ranks-higgs-two-extra}
\eeqa
Clearly, there is a second choice of fractional brane bounded by $X$, $Y$, which corresponds to 2, 3, 7, 12 in Figure \ref{fig:resolv1}(a), corresponding to 2, 3 in Figure \ref{fig:resolv1}(b). This corresponds to a fractional D3-branes with opposite 2-cycle homology charge. To keep track of this charge, we take into account the orientation of the paths, so the fractional branes we use in (\ref{ranks-higgs-two-extra}) correspond to increasing the ranks of the faces in the strip bounded by $X-Y$ (i.e. by $X$ and the orientation-reversed $Y$).

Consider now the addition of $K$ fractional antibranes on the second throat. At the level of the charges, this is equivalent to decreasing some of the ranks of suitable faces, especifically those bounded by $Y'-X'$ (keeping track of orientation, as explained above) in Figure \ref{fig:resolv2}. Namely, the rank assignments in the gauge theory corresponding to the second IR throat are
\beqa
& n_1=M_2-K \quad , \quad n_6=M_2+P_2-K \nonumber \\
& n_2=M_2 \quad , \quad n_{12}=M_2+P_2 \ .
\label{ranks-throat2-extra}
\eeqa
Moving up in the UV to the level of the theory before the Higgsing, the rank assignments change from (\ref{ranks-higgs-two}) by
\beqa
& & \Delta n_1=\Delta n_{5}= \Delta n_6=\Delta n_{8}=-K \ .
\eeqa
The fact that this variation is precisely opposite to that in (\ref{ranks-higgs-two-extra}) means that the combined set of two objects carries no charge. 

{\em At the level of the charges}, the above description amounts to imposing a different split of ranks in the Higgs mechanism, changing the vevs (\ref{vevs-complete1}) to
\beqa
&& \Phi_{81}=\Phi_{58}^T
=\begin{pmatrix} 0_{(M_1+P_1+K)\times (M_1+K)} & \cr & v_2\,  \id_{(M_2-K)\times (M_2-K)} \end{pmatrix}\nonumber \\[2pt]
&& \Phi_{27}=\Phi_{32}^T=\begin{pmatrix} 0_{(M_1+P_1)\times M_1} & \cr & v_2\,  \id_{M_2\times M_2} \end{pmatrix}
\nonumber\\[2pt]
&& \Phi_{61}=\Phi_{56}^T=\begin{pmatrix} v_1 \id_{(M_1+K)\times (M_1+K)}& \cr & 0_{ (M_2+P_2-K)\times (M_2-K)}  \end{pmatrix}\nonumber \\[2pt]
&&\Phi_{12,7}=\Phi_{3,12}^T=\begin{pmatrix} v_1 \id_{M_1\times M_1}& \cr & 0_{ (M_2+P_2)\times M_2}  \end{pmatrix} \ .
\label{vevs-complete}
\eeqa
The Higgsing by these vevs reproduces two decoupled gauge sectors corresponding to the UV of the two throats, with ranks modified by $\mathbb{Z}_2$ fractional brane charge. This reproduces the  first throat with $K$ extra fractional branes, and the second throat with  reduced rank groups
\beqa
SU(M_2-K)_1\times SU(M_2)_2\times SU(M_2+P_2-K)_6\times SU(M_2+P_2)_{12}  \ .
\eeqa
Actually, in analogy with \cite{Kachru:2002gs} (see also \cite{Argurio:2006ny,Argurio:2007qk}), this gauge sector should be regarded as providing a supersymmetric groundstate in a field theory in which the antibrane configuration should correspond to a metastable state (especifically, $K$ fractional antibranes in the throat defined by the $SU(M_2)^2\times SU(M_2+P_2)^2 $ theory, so that the total charges match). The energy associated to the susy breaking is suppressed by the RG cascade, compared with the energies at which the splitting of the throat occurs, so this justifies the approximation of describing the splitting as a mere Higgs mechanism at those scales.

\subsection{Hanany-Witten T-dual of axion monodromy}
 \label{sec:hw-tdual}
 
The appearance of axion monodromy upon the introduction of the D5-brane admits a simple intuitive description in terms of a T-dual Hanany Witten brane configuration  \cite{Hanany:1996ie}, which directly connects with a picture developed in \cite{McAllister:2008hb}.

Recall the description of the singularity (\ref{true-singu}), namely (removing the primes)
\beqa
xy=t^3\nonumber \\
zw=t^2 \ .
\label{not-simple-singu}
\eeqa
This equations describe the geometry as the superimposition of a $\mathbb{Z}_3$ and a $\mathbb{Z}_2$ orbifolds.
A T-duality along the $\mathbb{S}^1$ in $(x,y)$, defined by the orbit (\ref{orbit-one}) would lead to a configuration given by a $\mathbb{Z}_2$ orbifold of Figure \ref{uvcascade}, i.e. with the $\mathbb{Z}_3$ orbifold T-dualized into three NS and NS' branes, but with an explicit $\mathbb{Z}_2$ orbifold geometry  (the $\mathbb{Z}_2$ acting as a sign flip in the directions 4589), similar to those considered in \cite{Lykken:1997gy}. Hence this T-duality does not geometrize the B-field on the 2-cycle collapsed at the $\mathbb{Z}_2$. So we instead T-dualize along the $\mathbb{S}^1$ parametrized by $\beta$ in the orbit
\beqa
z\to e^{i\beta } z\quad , \quad w\to e^{-i\beta} w \ .
\eeqa
In this picture, the $\mathbb{Z}_2$ orbifold is geometrized in the T-dual into two NS- and two NS'-branes, in a $\mathbb{Z}_3$ orbifold geometry. The structure of NS5-branes is manifest in the fact that the locus of degeneration of the $\mathbb{S}^1$ in $(z,w)$ is $t=0$ (with multiplicity 2), which corresponds by the first equation to $xy=0$ (with multiplicity 2). This describes two kinds of objects, i.e. along $x=0$ or along $y=0$. 

The B-field of the 2-cycle collapsed at the $\mathbb{Z}_2$ orbifold singularity is geometrized as the relative distance between the two (NS, NS') pairs. The other relative brane separations correspond to B-fields on 2-cycles which actually disappear due to the complex deformations of the singularity. This can be seen explicitly, by following the action of the deformations in the Hanany-Witten T-dual. So let us deform the singularity (\ref{not-simple-singu}) to (c.f. (\ref{deform-simple}))
\beqa
&& xy=(t-\epsilon_1)(t-\epsilon_2)(t-\epsilon_3)\nonumber \\
&& zw=t^2 \ .
\eeqa
Performing the T-duality in this deformed geometry, we see that the degeneration locus of the $\mathbb{S}^1$ is $t=0$ (with multiplicity 2), which now corresponds to $xy={\rm const}$ (with multiplicity 2); this describes two copies of a unique kind of object, which is a recombination of the NS and NS'-brane. In other words, the complex deformation corresponds to shrinking the intervals within each (NS,NS') pair and combining the branes in the pair into a bound state. In the following we refer to this combined object as an NS5-brane (along the $t=0$).

\begin{figure}[htb]
\begin{center}
\vspace{.5cm}
\includegraphics[scale=.7]{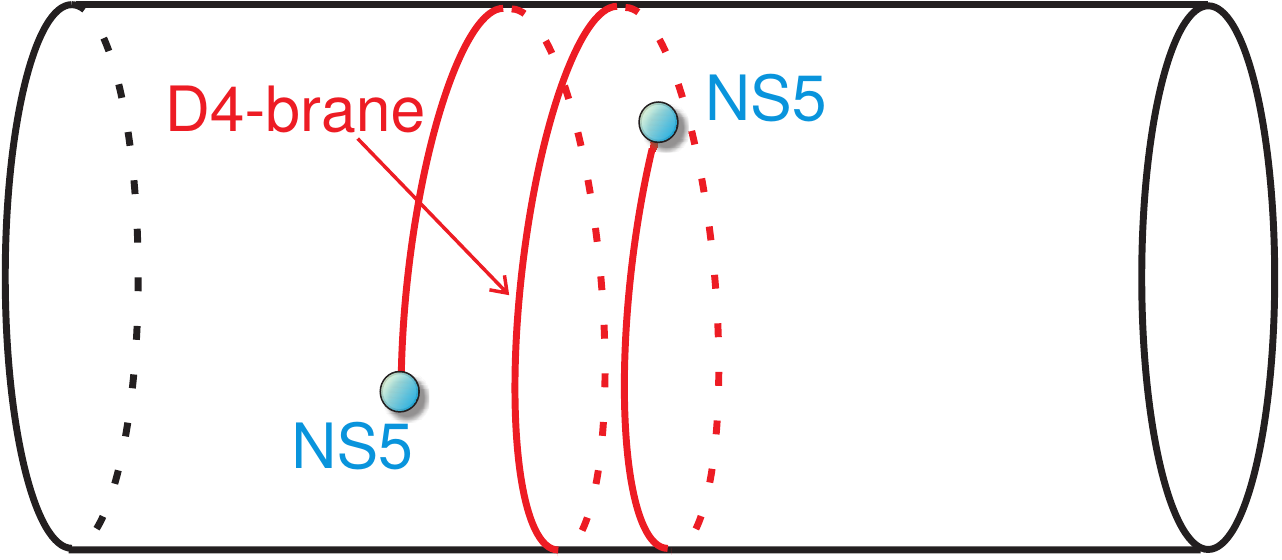}
\caption{\small T-dual configuration of the fractional D5-brane at the deformed singularity. The picture is precisely as in \cite{McAllister:2008hb}.}
\label{fig:windup}
\vspace*{.5cm}
\end{center}
\end{figure}

The B-field of the 2-cycle collapsed at the $\mathbb{Z}_2$ orbifold singularity corresponds to the surviving distance between the two NS5-branes. Also, the fractional D5-brane wrapping the collapsed 2-cycle corresponds to a D4-brane suspended along the interval between the NS5-branes. In this picture, the axion monodromy is manifest, and corresponds to the additional winding of the D4-branes when dragged by the relative motion of the two NS5-branes, see Figure \ref{fig:windup}, as described in \cite{McAllister:2008hb}. 

Actually, because the singularity contains a D5/anti-D5 pair located at different points of the curve of $\mathbb{C}^2/\mathbb{Z}_2$ singularities, the Hanany-Witten T-dual contains one anti-D4-brane stretched between the NS5-branes, in addition to the above mentioned D4-branes. The D4- and anti-D4-branes are located at different positions along the NS5-branes.

\subsection{Brane-antibrane backreaction}
\label{sec:back-cascade}

The holographic dual description can be used to address questions like the backreaction of the brane-antibrane system in the throat geometry.

There are two kinds of backreaction we can consider. The first is due to the presence of anti-D3-brane charge in the second throat. This bulk antibrane charge, in an otherwise supersymmetric throat, is completely similar to the anti-D3-branes in Klebanov-Strassler throats \cite{Kachru:2002gs}. These have been used in several applications \cite{Kachru:2003aw}, and are the subject of heated controversy concerning its backreacted solution in supergravity (see e.g. \cite{Bena:2009xk,Bena:2011hz,Bena:2011wh,Blaback:2012nf,Bena:2012bk,Bena:2013hr,Blaback:2014tfa,Bena:2014jaa,Danielsson:2014yga,Bena:2015kia}). We have nothing to add to the debate, except to mention that recent physical insights open the possibility of metastable regimes with antibranes \cite{Michel:2014lva}.

The novel feature about fivebrane-antibrane pairs in bifid throats is the existence of a backreaction on closed string fields associated to the homologous 2-cycle \cite{Conlon:2011qp}. In our example, these fields correspond to closed string twisted states at the $\mathbb{C}^2/\mathbb{Z}_2$ curve of singularities. On the geometrical side, one stack of fractional branes sources this field and leads to a log profile for it (see e.g. \cite{Karch:1998yv}), so it does not decrease as one moves away from the stack (the analysis in \cite{Conlon:2011qp} dealt with flat space or weakly curved geometries, the logs are still present in warped geometries \cite{Grana:2001xn}).

This behaviour is easily reproduced in the holographic field theory describing  the two IR throats. In fact, what follows is a simple generalization of what happens for fractional D3-branes at a $\mathbb{C}^2/\mathbb{Z}_2$ singularity (see e.g. \cite{Karch:1998yv}): the closed string twisted fields couple to the fractional brane by contributing to their gauge coupling constant, and the log dependence on the 2-plane transverse to the D3-branes is just the log dependence of the gauge coupling with the Coulomb branch parameter, controlled (at long distance / vev large compared with the strong dynamics scale) by the perturbative beta function of the corresponding ${\cal N}=2$ gauge theory.

We can reproduce the same analysis for the backreaction of the fractional branes in e.g. the first IR throat. 
Consider the theory with ranks (\ref{ranks-throat1-extra})  at the scale corresponding to the next-to-last step in the cascade. It has gauge group
\beqa
SU(P_1+K)_1\times SU(2P_1)_2\times SU(P_1)_3\times SU(2P_1+K)_8
\eeqa
and chiral bifundamentals $X_{12}$, $X_{21}$, $X_{23}$, $X_{32}$, $X_{38}$, $X_{83}$, $X_{81}$, $X_{18}$ and a superpotential clear from the dimer, but that we ignore. The non-perturbative dynamics of this theory can be analyzed directly, and reproduces (a number of fractional branes in) the deformed geometry. The result of interest can be obtained more easily as follows. There is a flat direction corresponding to giving vevs to $X_{81}$, $X_{18}$, which corresponds to moving the $K$ $\mathbb{Z}_2$ fractional branes away from the singular point, along the curve of $\mathbb{C}^2/\mathbb{Z}_2$ singularities, while the left-over $SU(P_1)^2\times SU(2P_1)^2$ theory generates the deformation of the throat as usual. The flat direction is actually the Coulomb branch of the $\mathbb{Z}_2$ fractional branes. Denoting by $\Lambda_i$ the dynamical scales of the gauge factors before taking the flat direction, and $\Lambda_i'$ the scales after integrating out the modes made massive by the vev $z$, the matching of scales gives
\beqa
\! \! \! \! \! \! \! \! \! \Lambda_1'{}^{-P_1}=\Lambda_1^{-P_1+2K} z^{-2K}\;\;\;\; , \;\;\;\;
\Lambda_2'{}^{4P_1}=\Lambda_2^{4P_1-K} z^{K} \hspace*{15pt}\nonumber \\
\Lambda_3'{}^{-P_1}=\Lambda_3^{-P_1-K} z^{K}\hspace*{28pt} , \;\;\;\;
\Lambda_8'{}^{4P_1}=\Lambda_8^{4P_1+2K} z^{-2K} \ . \nonumber
\eeqa
Relating the dynamical scales and the gauge couplings at some scale $\mu$
\beqa
\Lambda^{3N_c-N_f}=\mu^{3N_c-N_f} \exp \Big( \, \frac{1}{g_{\rm YM}^2(\mu)}+i\theta\, \Big)
\eeqa 
we have the following parametric dependence of the different gauge couplings (for gauge factors labelled by $i$)
\beqa
\frac{1}{g_{{\rm YM},i}^2(\mu)}+i\theta_i \,\sim\, K\log  z
\eeqa
This is the field theory description of the log dependence in the backreaction of fractional branes. The effect of these modifications in the RG flow that describes the throat is amenable to quantitative study, but on general grounds it does not spoil the geometric picture: the log is comparable to that generated by the fractional branes associated to the fluxes, but its coefficient is suppressed by the factor $K/P_i\ll 1$.

Moreover, the backreactions due to the brane-antibrane pair disappear as soon as both throats are combined into a single one, namely at the scale corresponding to the Higgsing separating the two IR throats. This is manifest in the field theory description of the introduction of the $\mathbb{Z}_2$ fractional  branes in terms of a specific choice of vevs. Above such scale, the vevs are negligible and fractional branes can be ignored.

Moreover, extending the discussion above we can estimate the backreaction effect in our holographic gauge theory description. For simplicity, we set the daughter throats symmetrical, and hence work with a hierarchy of scales $\Lambda\gg v\gg \Lambda'\equiv\Lambda_1=\Lambda_2$. In the singular limit of the original $\mathbb{Z}_3$ singularity, this corresponds to a hierarchy $N>M>P_i \equiv P_1=P_2$, where $N$ denotes effective D3-brane at the UV end of the parent throat, and we put $M$ and $P_i$ units of RR 3-form flux on the deformation A-cycles at the IR end of the parent throat which forms the UV of the daughter throats, and at the IR end of the daughter throats, respectively. Furthermore, we put $Q$ units of NSNS 3-form flux at the dual B-cycle of the parent throat and $Q'$ units of NSNS 3-form flux on the dual B-cycles of the daughter throats. We then get for the warp factors $\Lambda/\mp=e^{A}$ and $\Lambda'/\mp=e^{A'}$ at the bottom of the parent throat (and thus top of the daughter throats) and the bottom of the daughter throats, respectively, expressions reading~\cite{Giddings:2001yu}
\beq\label{eq:GKPwarp}
\Lambda \sim e^{-\frac{2\pi}{3}\frac{Q}{Mg_s}}\quad,\quad \Lambda' \sim e^{-\frac{2\pi}{3}\frac{Q'}{Pg_s}} \ .
\eeq
Now we need estimates for the scales $\Lambda,\Lambda'$. For this we use the results of the original geometric description of 5-brane axion monodromy~\cite{McAllister:2008hb}. Axion monodromy inflation arises now from the tension and the action of an NS5-brane wrapping a small resolution 2-cycle $\Sigma$ at the end of the curves of $\mathbb{C}^2/\mathbb{Z}_2$ singularities at the bottom (IR) of the daughter throats. Consider first the DBI action of a D5-brane
\beq
S_{D5}=\underbrace{\frac{1}{(2\pi)^5g_s\alpha'^3}}_{T_{\rm D5}}\int\limits_{M_4\times\Sigma} d^6\xi \sqrt{-\det (G+B_2)} \ .
\eeq
By S-duality the corresponding part the NS5-brane action reads
\beq
S_{NS5}=\underbrace{\frac{1}{(2\pi)^5g_s^2\alpha'^3}}_{T_{\rm NS5}}\int\limits_{M_4\times\Sigma} d^6\xi \sqrt{-\det (G+g_s^2C_2)} \ .
\eeq
If we denote the volume of the 2-cyle $\Sigma$ by $r^2\equiv vol(\Sigma)$, then in terms of the RR-axion $c\equiv \int_\Sigma C_2$ we have
\beq
\int\limits_{\Sigma} d^2y \sqrt{-\det (G+g_s^2C_2)}=\sqrt{r^4+g_s^2 c^2} \ .
\eeq
Hence, for large fields we get a linear inflationary axion potential, where we now include the overall warping of the NS5-brane energy density at the bottom of the daughter throat
\bea
V=e^{4A_{IR}^{daughter}}\frac{1}{(2\pi)^5g_s^2\alpha'^3} g_s \int_\Sigma C_2&\sim&\mp^4 \left(\frac{\Lambda'}{\mp}\right)^4\frac{g_s c}{4(2\pi)^3g_s\vol_E^2}\nonumber\\
&&\\
&\sim& \mp^4\, \frac{1}{8\pi \sqrt{g_s}\vol_E^{5/3}}\,\frac{\Lambda'^4}{\Lambda\mp^3}\,\frac{\phi_c}{\mp}\equiv \mu^4\,\frac{\phi_c}{\mp} \ .\nonumber
\eea
Here we used that $\alpha'\mp^2=\frac{2}{(2\pi)^7} \vol/g_s^2=\vol_E/(\pi\sqrt{g_s})$ and $\vol_E=g_s^{-3/2}/(2\pi)^6\vol\,,\,\vol=L^6$ denotes the 4D Einstein frame Calabi-Yau volume (in a suitably global version of our construction). 

Moreover, the curves of $\mathbb{C}^2/\mathbb{Z}_2$ singularities on which the $C_2$-axion is supported only reach up to the IR scale $\Lambda$ of the parent throat, which affects the definition of the canonically normalized inflaton field as in
\beq
\frac{\phi_c}{\mp}=\frac{\Lambda}{\mp}\frac{g_s c}{\sqrt{g_s}(2\pi)^2\vol_E^{1/3}} \ .
\eeq
Altogether, this produces the $\Lambda,\Lambda'$ dependence above. Imposing COBE normalization of the curvature perturbation power spectrum at the value $\phi_c=11\mp$ corresponding to 60 e-folds of slow-roll inflation yields the condition
\beq
\frac{\mu^4}{\mp^4}\simeq 2.2\times 10^{-10} \ .
\eeq
The resulting condition on the throat scales reads
\beq
\frac{\Lambda'^4}{\Lambda\mp^3}\sim 3.8\times10^{-6}\sqrt{\frac{g_s}{0.1}}\left(\frac{\vol_E}{100}\right)^{5/3} \ .
\eeq
Requiring the desired hierarchy $\Lambda>\Lambda'$ implies a lower bound on $\Lambda$ given by
\beq
\Lambda \gtrsim 0.016  \left(\frac{g_s}{0.1}\right)^{1/6}\left(\frac{\vol_E}{100}\right)^{5/9} \ .
\eeq
For equality the daughter throats would vanish into the parent throat as then $\Lambda=\Lambda'$.

An example is useful to give a feeling for the typical hierarchies achievable. If we take as typical values $g_s\sim 0.1$ and $\vol_E\sim 100$ then we can satisfy this condition e.g. by choosing $\Lambda\sim 0.3$ and $\Lambda'\sim 0.03$. To give an example, we can realize such a choice using eq.~\eqref{eq:GKPwarp} by turning on (non-compact) B-cycle NSNS flux quanta $Q=Q'$ and A-cycle RR flux quanta $M=17 Q$, $P_i=6 Q$ which satisfies the above constraint $M>P_i$.

If our axion inflaton potential above arises from $K$ NS5-branes and anti-NS5-branes wrapped on the small 2-cycles at the bottom of the daughter throats, then our above backreaction estimate expressed as ratio over the $P_i$ background fractional branes becomes
\beq
\frac{\Delta g_{{\rm YM},i}^{-2}(\mu)}{g_{{\rm YM}}^{-2}(\mu)}\sim \frac{K}{P_i}\log\frac{z}{\Lambda'}<\frac{K}{6Q}\log\frac{\Lambda}{\Lambda'}\simeq 2.3\times \frac{K}{6Q}\ll 1 \ .
\eeq
Here we used, that the top-to-bottom radial distance in the daughter throats $z$ is bounded by throat splitting VEV $v$ which in turn must sit below the IR scale of the parent throat $z<v<\Lambda$. In conclusion, the fractional size of the backreaction can be almost arbitrarily small, given that we can choose the NSNS flux $Q$ large subject only to tadpole bounds. 

We note here, that this is to be expected in a weakly coupled situation: the backreaction contribution denotes nothing else than the logarithmically scale-dependent 1-loop correction to the D4-brane world-volume gauge coupling in the Hanany-Witten T-dual 'brane-box' picture discussed in the previous section.

The above results also clearly tell us that for regimes with realistic scales the 5-brane-anti-brane backreaction is clearly subdominant to the backreaction driven by the 3-brane charge induced on the 5-branes by the wound-up axion~\cite{McAllister:2008hb,Flauger:2009ab}. The induced 3-brane charge is proportional to the axion winding number corresponding to a given canonical inflaton field displacement. This number of axion windings in turn scales inversely with the axion decay constant $f_a$. The decay constant arises from an integral over the homologous 2-cycle family reaching from one daughter throat into the other. This will suppress $f_a$ by the warp factor at the top of the daughter throats, that is, the scale $\Lambda$ at the bottom of the parent throat~\cite{McAllister:2008hb}. For a parametric estimate of this effect, see e.g.~\cite{Franco:2014hsa}, where the increased winding number is also shown to have a negligible impact on the tunneling rate despite the enhanced number of monodromy branches. Hence, warping the whole 2-cycle family setup as in our bifid setup does increase the amount of 3-brane charge build-up and its backreaction compared to the unwarped `snake' of~\cite{McAllister:2008hb}. However, imposing realistic scales constrains $\Lambda\gtrsim 0.02$, while in our example we have even $\Lambda\sim 0.3$. This serves to demonstrate that in phenomenologically viable setups the warping reduction of $f_a$ usually constitutes a rather mild effect. We leave a full computation of  $f_a$ in our bifid setup along the lines of~\cite{McAllister:2008hb,Franco:2014hsa} for the future.

\section{Conclusions}
\label{sec:conclusions}

The original models of axion monodromy inflation used a brane-antibrane pair of NS5-branes wrapped on the opposite-end minimum volume representatives of a homological family of 2-cycles which has to reach down into a bifurcated warped throat region. Control of backreaction issues requires this geometric structure to arise at the bottom of a warped parent throat, forming a `bifid' throat~\cite{Flauger:2009ab}. While this setup has its benefits like allowing forms of rigid moduli stabilization due to the localized nature of high-codimension branes sourcing the monodromy, a realization of the geometry was never done in a fully explicit local construction so far.

In this paper we described a very simple explicit local geometry realizing a bifid throat. $\mathbb{Z}_3$-orbifolding a conifold geometry provides a description of three warped conifolds with explicit holographic Klebanov-Strassler (KS) duals in terms of D3-branes at the toric singularities describing the three conifolds. Independently deforming them by small A-type 3-cycles provides then a manifest local construction of a bifid throat.  Its holographic dual description is given in terms of the fractional branes generating the different duality cascades, and the Higgsing  in the holographic picture corresponds to the splitting of the bottom of the parent throat into two independent daughter throats.

Further orbifolding this setup by another $\mathbb{Z}_2$ doubles this bifid throat into containing three pairs of deformed conifolds, which are connected via a pair of curves of $\mathbb{C}^2/\mathbb{Z}_2$-singularities reaching down into each daughter throat. Resolving them produces the homological 2-cycle family reaching down to the bottom of each daughter. We showed that the complex geometry describing this curve-pair of resolved $\mathbb{C}^2/\mathbb{Z}_2$-singularities reproduces the local resolved 2-conifold geometry of~\cite{Aganagic:2006ex}. As this appears here embedded in our full construction via D3-branes at the toric singularities of an orbifolded conifold, we effectively provided an ${\cal N}=1$ holographic dual for the setup in~\cite{Aganagic:2006ex}. Following~\cite{Aganagic:2006ex} further, the gauge theory description of 5-branes wrapped on the opposite-end 2-cycles reproduces the large-field linear axion potential known from 5-brane DBI action on the gravity side.

The simplicity of the orbifolded conifold geometry allows us also to identify explicitly the direction along which to perform a Hanany-Witten T-duality. This transforms the D5-brane axion monodromy system into the `brane box' picture of~\cite{McAllister:2008hb}. In this dual formulation there are two moving NS5-branes with the $B_2$-field on the 2-cycle replaced by the D4-brane stretching over multiple winding between the NS5-branes. Using this language we can also visualize the 5-brane backreaction effect as the logarithmically running 1-loop correction to the D4-brane worldvolume gauge coupling.

Having in place the explicit local construction, we can access the effect of backreaction from the NS5 brane-antibrane pair by looking at the holographic gauge theory description of small numbers of fractional 5- and anti-5-branes wrapped on the opposite-end 2-cycles, which captures the warping effect as well. We rediscover the logarithmic dependence found in~\cite{Conlon:2011qp} of the backreaction induced correction to the warp factors and hence scalar potential. Furthermore,  our holographic description  allows us to describe the warping neglected in~\cite{Conlon:2011qp}, so we can estimate the size of the logarithmic correction. 

We find that the warp factor hierarchies possible between the parent and daughter throats under the constraint of COBE normalization of the 5-brane monodromy inflation potential effectively constrain the sizes of the logarithm to be ${\cal O}(1)$ instead of being large. By  placing a small number $K$ of 5-brane-antibrane pairs into a given flux background generating the bifid throat,  we find that the logarithmic backreaction-induced correction is further suppressed for a choice of background fluxes providing large effective D3-charge compared to the $K$ 5-brane antibrane pairs.

Hence, our bifid throat construction with an explicit holographic dual provides us with an improved treatment of 5-brane-antibrane backreaction in the presence of warping. Beyond that, we may speculate that our setup may allow to study the effects of UV physics from compactification on 5-brane axion monodromy using similar holographic methods as applied to the holographic models of warped D3-brane inflation~\cite{Baumann:2008kq}.

\bigskip

\section*{Acknowledgments}

We thank Luis Ib\'a\~nez, Fernando Marchesano, Liam McAllister, Miguel Montero and Irene Valenzuela for useful discussions. AR and AU are partially supported by the grants  FPA2012-32828 from the MINECO, the ERC Advanced Grant SPLE under contract ERC-2012-ADG-20120216-320421 and the grant SEV-2012-0249 of the ``Centro de Excelencia Severo Ochoa" Programme. AW is supported by the Impuls und Vernetzungsfond of the Helmholtz Association of German Research Centers under grant HZ-NG-603.

\newpage



\begin{thebibliography}{99}

%
%
  
\bibitem{Ade:2015tva}
  P.~A.~R.~Ade {\it et al.}  [BICEP2 and Planck Collaborations],
  ``A Joint Analysis of BICEP2/Keck Array and Planck Data,''
  [arXiv:1502.00612 [astro-ph.CO]].
  
\bibitem{Creminelli:2015oda}
  P.~Creminelli, D.~L.~Nacir, M.~Simonovi\'{c}, G.~Trevisan and M.~Zaldarriaga,
  ``Detecting Primordial $B$-Modes after Planck,''
  arXiv:1502.01983 [astro-ph.CO].
  
\bibitem{Baumann:2014nda} 
  D.~Baumann and L.~McAllister,
 ``Inflation and String Theory,''
  arXiv:1404.2601 [hep-th].
  
\bibitem{Lyth:1996im} 
  D.~H.~Lyth,
 ``What would we learn by detecting a gravitational wave signal in the cosmic microwave background anisotropy?,''
  Phys.\ Rev.\ Lett.\  {\bf 78}, 1861 (1997)
  [hep-ph/9606387].
  
\bibitem{Kim:2004rp} 
  J.~E.~Kim, H.~P.~Nilles and M.~Peloso,
  ``Completing natural inflation,''
  JCAP {\bf 0501}, 005 (2005)
  [hep-ph/0409138].
  
\bibitem{Dimopoulos:2005ac} 
  S.~Dimopoulos, S.~Kachru, J.~McGreevy and J.~G.~Wacker,
 ``N-flation,''
  JCAP {\bf 0808}, 003 (2008)
  [hep-th/0507205].

\bibitem{Grimm:2007hs} 
  T.~W.~Grimm,
 ``Axion inflation in type II string theory,''
  Phys.\ Rev.\ D {\bf 77}, 126007 (2008)
  [arXiv:0710.3883 [hep-th]].

\bibitem{Berg:2009tg} 
  M.~Berg, E.~Pajer and S.~Sjors,
 ``Dante's Inferno,''
  Phys.\ Rev.\ D {\bf 81}, 103535 (2010)
  [arXiv:0912.1341 [hep-th]].
  
\bibitem{Ben-Dayan:2014lca}
  I.~Ben-Dayan, F.~G.~Pedro and A.~Westphal,
  ``Towards Natural Inflation in String Theory,''
  arXiv:1407.2562 [hep-th].
  
\bibitem{Ben-Dayan:2014zsa}
  I.~Ben-Dayan, F.~G.~Pedro and A.~Westphal,
  ``Hierarchical Axion Inflation,''
  Phys.\ Rev.\ Lett.\  {\bf 113} (2014) 26,  261301
  [arXiv:1404.7773 [hep-th]].

\bibitem{Silverstein:2008sg}
  E.~Silverstein and A.~Westphal,
  ``Monodromy in the CMB: Gravity Waves and String Inflation,''
  Phys.\ Rev.\ D {\bf 78} (2008) 106003
  [arXiv:0803.3085 [hep-th]].
 
  
\bibitem{McAllister:2008hb}
  L.~McAllister, E.~Silverstein and A.~Westphal,
  ``Gravity Waves and Linear Inflation from Axion Monodromy,''
  Phys.\ Rev.\ D {\bf 82} (2010) 046003
  [arXiv:0808.0706 [hep-th]].
  
\bibitem{Marchesano:2014mla} 
  F.~Marchesano, G.~Shiu and A.~M.~Uranga,
  ``F-term Axion Monodromy Inflation,''
  JHEP {\bf 1409}, 184 (2014)
  [arXiv:1404.3040 [hep-th]].

\bibitem{Blumenhagen:2014gta}
  R.~Blumenhagen and E.~Plauschinn,
  ``Towards Universal Axion Inflation and Reheating in String Theory,''
  Phys.\ Lett.\ B {\bf 736} (2014) 482
  [arXiv:1404.3542 [hep-th]].
  
\bibitem{Hebecker:2014eua}
  A.~Hebecker, S.~C.~Kraus and L.~T.~Witkowski,
  ``D7-Brane Chaotic Inflation,''
  Phys.\ Lett.\ B {\bf 737} (2014) 16
  [arXiv:1404.3711 [hep-th]].
  
\bibitem{McAllister:2014mpa}
  L.~McAllister, E.~Silverstein, A.~Westphal and T.~Wrase,
  ``The Powers of Monodromy,''
  JHEP {\bf 1409} (2014) 123
  [arXiv:1405.3652 [hep-th]].
  
\bibitem{Garcia-Etxebarria:2014wla} 
  I.~García-Etxebarria, T.~W.~Grimm and I.~Valenzuela,
  ``Special Points of Inflation in Flux Compactifications,''
  arXiv:1412.5537 [hep-th].
  
\bibitem{Ibanez:2014kia}
  L.~E.~Ibanez and I.~Valenzuela,
  ``The inflaton as an MSSM Higgs and open string modulus monodromy inflation,''
  Phys.\ Lett.\ B {\bf 736} (2014) 226
  [arXiv:1404.5235 [hep-th]].
  
\bibitem{Franco:2014hsa}
  S.~Franco, D.~Galloni, A.~Retolaza and A.~Uranga,
  ``Axion Monodromy Inflation on Warped Throats,''
  JHEP {\bf 1502} (2015) 086
  [arXiv:1405.7044 [hep-th]].
  
\bibitem{Blumenhagen:2014nba}
  R.~Blumenhagen, D.~Herschmann and E.~Plauschinn,
  ``The Challenge of Realizing F-term Axion Monodromy Inflation in String Theory,''
  JHEP {\bf 1501} (2015) 007
  [arXiv:1409.7075 [hep-th]].
  
\bibitem{Hebecker:2014kva}
  A.~Hebecker, P.~Mangat, F.~Rompineve and L.~T.~Witkowski,
  ``Tuning and Backreaction in F-term Axion Monodromy Inflation,''
  arXiv:1411.2032 [hep-th].
  
\bibitem{Ibanez:2014swa}
  L.~E.~Ibanez, F.~Marchesano and I.~Valenzuela,
  ``Higgs-otic Inflation and String Theory,''
  JHEP {\bf 1501} (2015) 128
  [arXiv:1411.5380 [hep-th]].
  
\bibitem{Kaloper:2008fb}
  N.~Kaloper and L.~Sorbo,
  ``A Natural Framework for Chaotic Inflation,''
  Phys.\ Rev.\ Lett.\  {\bf 102} (2009) 121301
  [arXiv:0811.1989 [hep-th]].

\bibitem{Kaloper:2011jz}
  N.~Kaloper, A.~Lawrence and L.~Sorbo,
  ``An Ignoble Approach to Large Field Inflation,''
  JCAP {\bf 1103} (2011) 023
  [arXiv:1101.0026 [hep-th]].
  
  \bibitem{Kaloper:2014zba} 
  N.~Kaloper and A.~Lawrence,
   ``Natural Chaotic Inflation and UV Sensitivity,''
  arXiv:1404.2912 [hep-th].
  
\bibitem{Banks:2003sx}
  T.~Banks, M.~Dine, P.~J.~Fox and E.~Gorbatov,
  ``On the possibility of large axion decay constants,''
  JCAP {\bf 0306} (2003) 001
  [hep-th/0303252].
\bibitem{ArkaniHamed:2006dz}
  N.~Arkani-Hamed, L.~Motl, A.~Nicolis and C.~Vafa,
  ``The String landscape, black holes and gravity as the weakest force,''
  JHEP {\bf 0706} (2007) 060
  [hep-th/0601001].

\bibitem{Rudelius:2015xta}
  T.~Rudelius,
  ``Constraints on Axion Inflation from the Weak Gravity Conjecture,''
  arXiv:1503.00795 [hep-th].
  
\bibitem{Montero:2015ofa}
  M.~Montero, A.~M.~Uranga and I.~Valenzuela,
  ``Transplanckian axions !?,''
  arXiv:1503.03886 [hep-th].
  
\bibitem{Brown:2015iha}
  J.~Brown, W.~Cottrell, G.~Shiu and P.~Soler,
  ``Fencing in the Swampland: Quantum Gravity Constraints on Large Field Inflation,''
  arXiv:1503.04783 [hep-th].
  
\bibitem{Bachlechner:2015qja}
  T.~C.~Bachlechner, C.~Long and L.~McAllister,
  ``Planckian Axions and the Weak Gravity Conjecture,''
  arXiv:1503.07853 [hep-th].
  
\bibitem{Hebecker:2015rya}
  A.~Hebecker, P.~Mangat, F.~Rompineve and L.~T.~Witkowski,
  ``Winding out of the Swamp: Evading the Weak Gravity Conjecture with F-term Winding Inflation?,''
  arXiv:1503.07912 [hep-th].
  
\bibitem{Brown:2015lia}
  J.~Brown, W.~Cottrell, G.~Shiu and P.~Soler,
  ``On Axionic Field Ranges, Loopholes and the Weak Gravity Conjecture,''
  arXiv:1504.00659 [hep-th].
  
\bibitem{Palti:2014kza}
  E.~Palti and T.~Weigand,
  ``Towards large r from [p, q]-inflation,''
  JHEP {\bf 1404} (2014) 155
  [arXiv:1403.7507 [hep-th]].
  
\bibitem{Flauger:2009ab} 
  R.~Flauger, L.~McAllister, E.~Pajer, A.~Westphal and G.~Xu,
   ``Oscillations in the CMB from Axion Monodromy Inflation,''
  JCAP {\bf 1006}, 009 (2010)
  [arXiv:0907.2916 [hep-th]].
   
\bibitem{Conlon:2011qp}
  J.~P.~Conlon,
  ``Brane-Antibrane Backreaction in Axion Monodromy Inflation,''
  JCAP {\bf 1201} (2012) 033
  [arXiv:1110.6454 [hep-th]].
 
\bibitem{Klebanov:2000hb}
  I.~R.~Klebanov and M.~J.~Strassler,
  ``Supergravity and a confining gauge theory: Duality cascades and chi SB resolution of naked singularities,''
  JHEP {\bf 0008} (2000) 052
  [hep-th/0007191].
  
\bibitem{Aganagic:2006ex}
  M.~Aganagic, C.~Beem, J.~Seo and C.~Vafa,
  ``Geometrically induced metastability and holography,''
  Nucl.\ Phys.\  B {\bf 789}, 382 (2008)
  [arXiv:hep-th/0610249].
    
\bibitem{Franco:2005fd}
  S.~Franco, A.~Hanany and A.~M.~Uranga,
  ``Multi-flux warped throats and cascading gauge theories,''
  JHEP {\bf 0509} (2005) 028
  [hep-th/0502113].
  
\bibitem{Franco:2005rj} 
  S.~Franco, A.~Hanany, K.~D.~Kennaway, D.~Vegh and B.~Wecht,
  ``Brane dimers and quiver gauge theories,''
  JHEP {\bf 0601}, 096 (2006)
  [hep-th/0504110].

\bibitem{Kennaway:2007tq} 
  K.~D.~Kennaway,
  ``Brane Tilings,''
  Int.\ J.\ Mod.\ Phys.\ A {\bf 22}, 2977 (2007)
  [arXiv:0706.1660 [hep-th]].
  
\bibitem{Gukov:1999ya}
  S.~Gukov, C.~Vafa and E.~Witten,
  ``CFT's from Calabi-Yau four folds,''
  Nucl.\ Phys.\ B {\bf 584} (2000) 69
   [Erratum-ibid.\ B {\bf 608} (2001) 477]
  [hep-th/9906070].
  
\bibitem{Hanany:1996ie}
  A.~Hanany and E.~Witten,
  ``Type IIB superstrings, BPS monopoles, and three-dimensional gauge dynamics,''
  Nucl.\ Phys.\ B {\bf 492} (1997) 152
  [hep-th/9611230].
  
\bibitem{Uranga:1998vf}
  A.~M.~Uranga,
  ``Brane configurations for branes at conifolds,''
  JHEP {\bf 9901} (1999) 022
  [hep-th/9811004].
    
\bibitem{GarciaEtxebarria:2006aq}
  I.~Garcia-Etxebarria, F.~Saad and A.~M.~Uranga,
  ``Quiver gauge theories at resolved and deformed singularities using dimers,''
  JHEP {\bf 0606} (2006) 055
  [hep-th/0603108].
  
  
\bibitem{Kachru:2002gs}
  S.~Kachru, J.~Pearson and H.~L.~Verlinde,
  ``Brane / flux annihilation and the string dual of a nonsupersymmetric field theory,''
  JHEP {\bf 0206} (2002) 021
  [hep-th/0112197].
  
\bibitem{Argurio:2006ny}
  R.~Argurio, M.~Bertolini, S.~Franco and S.~Kachru,
  ``Gauge/gravity duality and meta-stable dynamical supersymmetry breaking,''
  JHEP {\bf 0701} (2007) 083
  [hep-th/0610212].

\bibitem{Argurio:2007qk}
  R.~Argurio, M.~Bertolini, S.~Franco and S.~Kachru,
  ``Meta-stable vacua and D-branes at the conifold,''
  JHEP {\bf 0706} (2007) 017
   [AIP Conf.\ Proc.\  {\bf 1031} (2008) 94]
  [hep-th/0703236].
  
    
\bibitem{Lykken:1997gy}
  J.~D.~Lykken, E.~Poppitz and S.~P.~Trivedi,
  ``Chiral gauge theories from D-branes,''
  Phys.\ Lett.\ B {\bf 416} (1998) 286
  [hep-th/9708134].
  
\bibitem{Kachru:2003aw}
  S.~Kachru, R.~Kallosh, A.~D.~Linde and S.~P.~Trivedi,
  ``De Sitter vacua in string theory,''
  Phys.\ Rev.\ D {\bf 68} (2003) 046005
  [hep-th/0301240].
  
\bibitem{Bena:2009xk}
  I.~Bena, M.~Grana and N.~Halmagyi,
  ``On the Existence of Meta-stable Vacua in Klebanov-Strassler,''
  JHEP {\bf 1009} (2010) 087
  [arXiv:0912.3519 [hep-th]].

\bibitem{Bena:2011hz}
  I.~Bena, G.~Giecold, M.~Grana, N.~Halmagyi and S.~Massai,
  ``On Metastable Vacua and the Warped Deformed Conifold: Analytic Results,''
  Class.\ Quant.\ Grav.\  {\bf 30} (2013) 015003
  [arXiv:1102.2403 [hep-th]].
 
\bibitem{Bena:2011wh}
  I.~Bena, G.~Giecold, M.~Grana, N.~Halmagyi and S.~Massai,
  ``The backreaction of anti-D3 branes on the Klebanov-Strassler geometry,''
  JHEP {\bf 1306} (2013) 060
  [arXiv:1106.6165 [hep-th]].
  
\bibitem{Bena:2012bk}
  I.~Bena, M.~Grana, S.~Kuperstein and S.~Massai,
  ``Anti-D3 Branes: Singular to the bitter end,''
  Phys.\ Rev.\ D {\bf 87} (2013) 10,  106010
  [arXiv:1206.6369 [hep-th]].
  
\bibitem{Bena:2014jaa}
  I.~Bena, M.~Grana, S.~Kuperstein and S.~Massai,
  ``Giant Tachyons in the Landscape,''
  arXiv:1410.7776 [hep-th].
   
\bibitem{Blaback:2012nf}
  J.~Blaback, U.~H.~Danielsson and T.~Van Riet,
  ``Resolving anti-brane singularities through time-dependence,''
  JHEP {\bf 1302} (2013) 061
  [arXiv:1202.1132 [hep-th]].
  
\bibitem{Bena:2013hr}
  I.~Bena, J.~Blaback, U.~H.~Danielsson and T.~Van Riet,
  ``Antibranes cannot become black,''
  Phys.\ Rev.\ D {\bf 87} (2013) 10,  104023
  [arXiv:1301.7071 [hep-th]].
  
\bibitem{Blaback:2014tfa}
  J.~Blaback, U.~H.~Danielsson, D.~Junghans, T.~Van Riet and S.~C.~Vargas,
  ``Localised anti-branes in non-compact throats at zero and finite $T$,''
  JHEP {\bf 1502} (2015) 018
  [arXiv:1409.0534 [hep-th]].
  
\bibitem{Danielsson:2014yga}
  U.~H.~Danielsson and T.~Van Riet,
  ``Fatal attraction: more on decaying anti-branes,''
  arXiv:1410.8476 [hep-th].
  
\bibitem{Bena:2015kia}
  I.~Bena and S.~Kuperstein,
  ``Brane polarization is no cure for tachyons,''
  arXiv:1504.00656 [hep-th].
    
\bibitem{Michel:2014lva}
  B.~Michel, E.~Mintun, J.~Polchinski, A.~Puhm and P.~Saad,
  ``Remarks on brane and antibrane dynamics,''
  arXiv:1412.5702 [hep-th].

\bibitem{Karch:1998yv}
  A.~Karch, D.~Lust and D.~J.~Smith,
  ``Equivalence of geometric engineering and Hanany-Witten via fractional branes,''
  Nucl.\ Phys.\ B {\bf 533} (1998) 348
  [hep-th/9803232].

\bibitem{Grana:2001xn}
  M.~Grana and J.~Polchinski,
  ``Gauge / gravity duals with holomorphic dilaton,''
  Phys.\ Rev.\ D {\bf 65} (2002) 126005
  [hep-th/0106014].
  
\bibitem{Giddings:2001yu} 
  S.~B.~Giddings, S.~Kachru and J.~Polchinski,
  ``Hierarchies from fluxes in string compactifications,''
  Phys.\ Rev.\ D {\bf 66}, 106006 (2002)
  [hep-th/0105097].
  
\bibitem{Baumann:2008kq}
  D.~Baumann, A.~Dymarsky, S.~Kachru, I.~R.~Klebanov and L.~McAllister,
  ``Holographic Systematics of D-brane Inflation,''
  JHEP {\bf 0903} (2009) 093
  [arXiv:0808.2811 [hep-th]].

\end{thebibliography}
\end{document}